\documentclass[11pt]{article}
\pdfoutput=1
\usepackage{jcappub}
\usepackage{graphicx}

\usepackage[dvipsnames]{xcolor}
\usepackage{xspace}

\usepackage[normalem]{ulem}

\newcommand{\overbar}[1]{\mkern 1.5mu\overline{\mkern-3.5mu#1\mkern-1.5mu}\mkern 1.5mu}

\newcommand{\threej}[6]{ \begin{pmatrix}
    #1 & #2 & #3 \\
    #4 & #5 & #6 
   \end{pmatrix}}
  
 \newcommand{\sixj}[6]{ \begin{Bmatrix}
    #1 & #2 & #3 \\
    #4 & #5 & #6 
   \end{Bmatrix}}

\newcommand{\ib}{\mathrm{I}\!B}
\newcommand{\ic}{\mathrm{IC}}

\newcommand{\fsky}{f_\mathrm{sky}}
\newcommand{\lmin}{\ell_\mathrm{min}}
\newcommand{\lmax}{\ell_\mathrm{max}}
\newcommand{\nside}{N_\mathrm{side}}

\newcommand{\lwmin}{\ell^\mathrm{min}_w}
\newcommand{\lwmax}{\ell^\mathrm{max}_w}
\newcommand{\euclid}{\textit{Euclid}}
\title{The integrated angular bispectrum of weak lensing}
\author[a,b]{Gabriel  Jung,}
\author[c]{Toshiya Namikawa,}
\author[a,b]{Michele Liguori,}
\author[d]{Dipak Munshi,}
\author[e]{Alan Heavens}

\affiliation[a]{Dipartimento di Fisica e Astronomia "G. Galilei", Universit{\`a} degli Studi di Padova,\\ Via Marzolo 8, 35131 Padova, Italy}
\affiliation[b]{INFN, Sezione di Padova, via Marzolo 8, I-35131, Padova, Italy}
\affiliation[c]{Department of Applied Mathematics and Theoretical Physics, University of Cambridge, Wilberforce Road, Cambridge CB3 0WA, U.K.}
\affiliation[d]{Mullard Space Science Laboratory,  \\ University College London, Holmbury St Mary,\\ Dorking, Surrey RH5 6NT, U.K.}
\affiliation[e]{Imperial Centre for Inference and Cosmology (ICIC), Imperial College, London, SW7 2AZ, U.K.}

\emailAdd{gabriel.jung@pd.infn.it}


\arxivnumber{2102.05521}

\abstract{We investigate three-point statistics in weak lensing convergence, through the integrated bispectrum.  This statistic involves measuring power spectra in patches, and is thus easy to measure, and avoids the complexity of estimating the very large number of possible bispectrum configurations.  The integrated bispectrum principally probes the squeezed limit of the bispectrum.  To be useful as a set of summary statistics, accurate theoretical predictions of the signal are required, and, assuming Gaussian sampling distributions, the covariance matrix.  In this paper, we investigate through simulations how accurate are theoretical formulae for both the integrated bispectrum and its covariance, finding that there a small inaccuracies in the theoretical signal, and more serious deviations in the covariance matrix, which may need to be estimated using simulations.}

\begin{document}

\maketitle
\flushbottom

\section{Introduction}

Weak gravitational lensing is a potentially powerful probe of cosmology as the observable signatures are readily connected to fundamental theory.  Recent studies have put constraints on the amplitude of matter fluctuations \citep{Hikage2019,Troxel2019,Asgari2020}, and future prospects with the Euclid satellite \citep{Laureijs2011} and the Legacy Survey of Space and Time (LSST)\footnote{\url{https://www.lsst.org}} with the Vera Rubin Observatory promise precise measurements.  The principal physical effects are purely gravitational, so are dependent on the distribution of matter in the Universe, which is easier to predict in theoretical models than the distribution of galaxies.  The traditional statistics to use for comparison of theory with data are two-point statistics - correlation functions or power spectra.  These can be sufficient statistics provided that the field under consideration is a random Gaussian field, but the non-linear growth of structure by gravitational instability means that the field is non-Gaussian, and the two-point statistics do not capture all of the cosmological information contained in the field. Thus, to fully exploit the data requires going beyond the power spectrum to probe also the non-Gaussian properties of the field.  There are several ways to do this, from sophisticated Bayesian forward-modelling techniques, which incorporate a gravity model \citep{Porqueres2021} and which apply the likelihood at the field level, likelihood-free inference \citep{Jeffrey2021}, or by analysing higher-order summary statistics. The addition of the bispectrum to the power spectrum can lead to significant reduction in errors \citep{Takada2004,Kayo2013,Rizzato2019} and better control of systematics \citep{Pyne2020}.  

There are alternative approaches to including non-Gaussian information, such as with skew-spectra \citep{Munshi2010,Munshi2020c} and Minkowski functionals \cite{Munshi2020b}, both of which have been developed for weak lensing.

For higher-order statistics, there are several challenges, one of which is the very large number of three-point functions (bispectra, in harmonic space) that can be considered.  In addition to this is the formidable challenge of making accurate theoretical predictions for these statistics \citep{Bernardeau2012} and to compute their sampling distribution. In this paper, in order to address the first complication, we consider the integrated bispectrum, which involves computing only power spectra, in patches on the sky, but which probes the squeezed limit of the bispectrum. This approach has been first proposed in the context of Large Scale Structure \cite{Chiang:2014oga} and has a wide range of applications like galaxy clustering, 21cm and weak lensing studies \cite{Chiang:2015eza,Chiang:2015pwa,Giri:2018dln,Munshi2017,Munshi2020a}. Here we consider its extension to 2D random fields on the sphere developed in \cite{Jung:2020zne} for CMB non-Gaussianity (NG) analyses. For comparison and validation purposes, we also study the binned bispectrum approach \cite{Bucher:2009nm, Bucher:2015ura, Munshi:2019csw}, where the extremely large number of modes is reduced by imposing a binning in harmonic space. We assume that these statistics have a Gaussian distribution, and compute their covariance in the limit of weak non-Gaussianity.  Finally, in this first paper, we consider the spin-0 convergence field, rather than the spin-2 cosmic shear field, as this field is easier to deal with, whilst still incorporating many of the same challenges as the cosmic shear field, which is the usual field studied with weak lensing.  Cosmic shear and the inclusion of NG contributions in the covariance will be the subjects of a future paper.  The magnification field can be probed with data, for example from size or flux measurements \citep{Alsing2015,Duncan2014,Duncan2016}, but our main focus here is the accuracy with which the integrated bispectrum may be predicted theoretically.

The outline of the paper is as follows: in section \ref{sec:weak-lensing} we review the bispectrum for weak lensing; in section \ref{sec:integrated-bispectrum} we introduce the estimator for the integrated bispectrum; in section \ref{sec:simulations} we present results from simulations; in section \ref{sec:binned-bispectrum} we validate our results by comparing them to the binned bispectrum statistic; in section \ref{sec:covariance} we study the covariance of the estimator, and in section \ref{sec:conclusion} we present our conclusions.

\section{The weak lensing convergence bispectrum}
\label{sec:weak-lensing}

In the Born approximation, the weak lensing convergence field is a weighted integral of the matter density contrast $\delta$ between the source plane (at comoving distance $r_s$, or redshift $z_s$) and the observer:
\begin{equation}
    \label{eq:convergence}
    \kappa(\hat{\Omega}, r_s) =  \int_0^{r_s} dr \, \omega(r, r_s)\delta(\hat{\Omega}, r)\,,
\end{equation}
where $\hat{\Omega}$ is the angular position on the celestial sphere, and $r$ is the comoving distance. The weights $\omega(r,r_s)$ are given by
\begin{equation}
\label{eq:convergence-weight}
    \omega(r, r_s)  = \frac{3\Omega_M}{2}\frac{H_0^ 2}{c^2}\frac{d_A(r)d_A(r-r_s)}{a(r)d_A(r_s)}\,,
\end{equation}
where $d_A(r)$ is the comoving angular diameter distance, $a(r)$ is the scale factor, and $\Omega_M$, $H_0$, $c$ are the cosmological matter density parameter, the Hubble constant and the speed of light respectively.   For reviews of weak lensing, see for example \citep{Bartelmann2001,Munshi2008,Kilbinger2015}.

In this paper, we are mainly interested in the three-point correlator of the convergence field, or in harmonic space, of its harmonic coefficients denoted as $\kappa_{\ell m}$. Under the assumption of statistical isotropy, it is fully described by the following angle-averaged bispectrum
\begin{equation}
    \label{eq:bispectrum}
    \begin{split}
    B_{\ell_1 \ell_2 \ell_3} &=h_{\ell_1 \ell_2 \ell_3} \sum\limits_{m_1 m_2 m_3} 
     \threej{\ell_1}{\ell_2}{\ell_3}{m_1}{m_2}{m_3} 
     \langle \kappa_{\ell_1 m_1} \kappa_{\ell_2 m_2} \kappa_{\ell_2 m_2}\rangle \\
     &= \left\langle \int d^2\hat\Omega \,
     \kappa_{\ell_1} (\hat\Omega) \kappa_{\ell_2} (\hat\Omega)\kappa_{\ell_3} (\hat\Omega)\right\rangle
     \,,
    \end{split}
\end{equation} 
where the matrix is a Wigner $3j$-symbol, $\kappa_{\ell}$ are maximally filtered maps given by \begin{equation}
    \kappa_\ell = \sum\limits_{m=-\ell}^{\ell}\kappa_{\ell m}Y_{\ell m}\,
\end{equation}
 and the geometrical factor $h_{\ell_1 \ell_2 \ell_3}$ is defined by
\begin{equation}
    \label{eq:h-factor}
    h_{\ell_1 \ell_2 \ell_3}
    \equiv
    \sqrt{\frac{(2\ell_1+1)(2\ell_2+1)(2\ell_3+1)}{4\pi}}
    \threej{\ell_1}{\ell_2}{\ell_3}{0}{0}{0}\,.
\end{equation}
In the literature, the reduced bispectrum $b_{\ell_1 \ell_2 \ell_3} = B_{\ell_1 \ell_2 \ell_3} / h_{\ell_1 \ell_2 \ell_3}^2$ is also often used. 

In the Limber approximation, the convergence angle-averaged bispectrum can be written in terms of the matter bispectrum $B_\delta(k_1, k_2, k_3)$, where  $k_i$ are comoving wavenumbers, as
\begin{equation}
    \label{eq:convergence-bispectrum}
    B_{\ell_1 \ell_2 \ell_3} =  h_{\ell_1 \ell_2 \ell_3}^2\int_0^{r_s} dr \, \frac{\omega(r, r_s)^3}{d_A^4(r_s)}B_\delta\left(\frac{\ell_1}{d_A(r)}, \frac{\ell_2}{d_A(r)}, \frac{\ell_3}{d_A(r)};r\right)\,.
\end{equation}
To compute this, we use the fitting function developed in  \cite{Takahashi:2019hth}. One can also includes the post-Born correction \cite{Pratten:2016dsm} which becomes necessary at high redshift (see section \ref{sec:cmb-lensing}) but has a small effect at low redshift (see section \ref{sec:ideal-simulations}). 
\section{The integrated angular bispectrum estimator}
\label{sec:integrated-bispectrum}

In this paper, we study the weak lensing convergence bispectrum with a simple method which does not require measurement of anything more complicated than power spectra (thus no three-point correlators). It is based on three relatively simple steps: separate the celestial sphere into many equal-sized patches, determine the power spectrum (small-scale fluctuations) and the average value (large-scale mode) in each patch and compute their patch-by-patch correlation averaged over the sky. The result is called the integrated bispectrum and is by construction sensitive to the correlations between small-scale and large-scale effects, like the squeezed limit of the bispectrum (one multipole much smaller than the other two).

\begin{figure}
    \centering
    \includegraphics[width=0.99\linewidth]{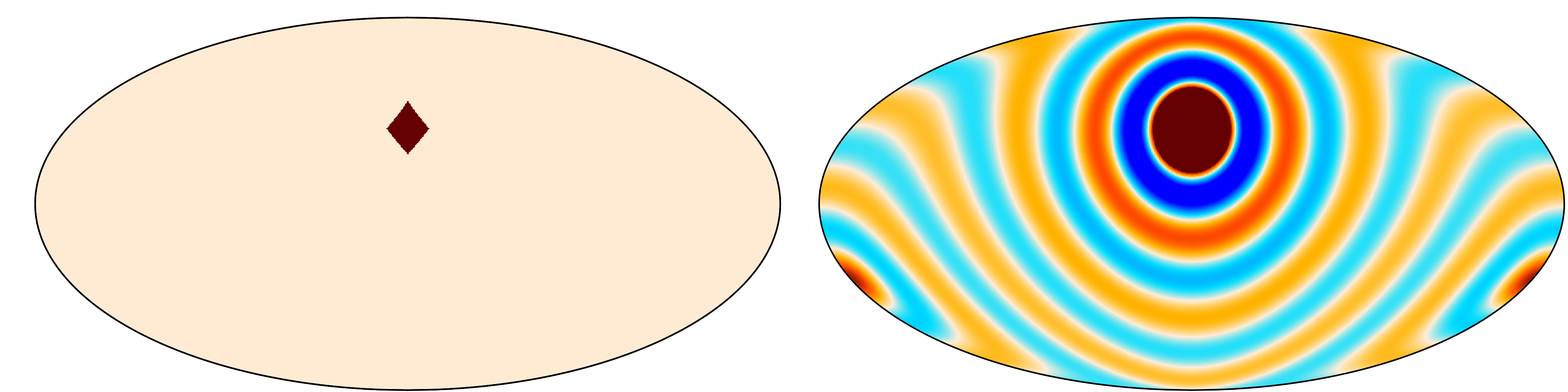}
    \includegraphics[width=0.5\linewidth]{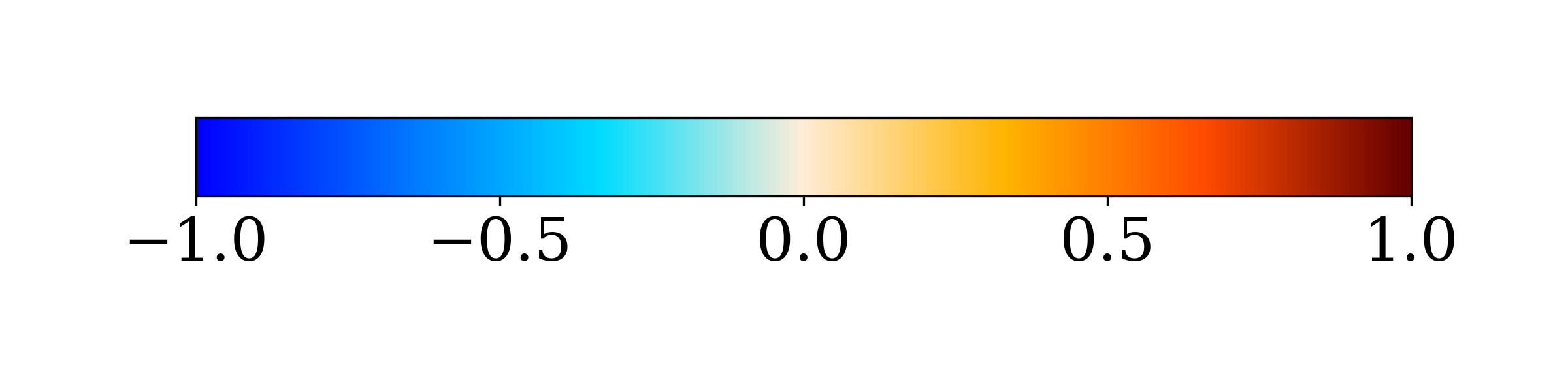}
    \caption{The two types of patches used in this paper. On the left, one HEALPix patch (among a set of 192), and on the right a step function patch centred at the same position and defined by eq.\ \eqref{eq:azimuth-patch} with $\lwmin = 0$ and $\lwmax = 10$.}
    \label{fig:patches}  
\end{figure}

Implementing this method first requires the characteristics of the patches (size, shape, number) to be specified, and then sky realizations $W_\mathrm{patch}(\hat{\Omega})$ produced, where the index `$\mathrm{patch}$' denotes the exact patch considered in the full set (see figure \ref{fig:patches} for examples). Then, one by one, these patch maps, are applied as masks to the observational data $\kappa^\mathrm{obs}(\Omega)$. For each resulting map, we only have to compute two simple quantities; its power spectrum $C_{\ell,\mathrm{patch}}^\mathrm{obs}$, which is called the position-dependent power spectrum in the literature, and its average value $\overbar{\kappa}_\mathrm{patch}^\mathrm{obs}$. The product of these two quantities is finally averaged over all the patches to obtain the integrated angular bispectrum estimator:
\begin{equation}
    \label{eq:ibisp-estimator}
    \ib_\ell^\mathrm{obs} = \frac{1}{N_\mathrm{patch}}\sum\limits_{\mathrm{patch}}
        \overbar{\kappa}_\mathrm{patch}^\mathrm{obs} C_{\ell,\mathrm{patch}}^\mathrm{obs}\,,
\end{equation}
where $N_\mathrm{patch}$ is the number of patches used to divide the sky. 
For example, a way to separate the sky into a set of equal-sized patches is to use the standard HEALPix\footnote{\url{http://healpix.sourceforge.net}} pixelization \cite{Gorski:2004by}: starting from a given data map, one degrades it to lower resolution and defines a patch by fixing every pixel of the low resolution map to zero, except one. The power spectrum in the chosen patch is then computed at high resolution. Repeating this process for each low-resolution pixel gives a set of patches that covers uniformly the full sky. 

When using observational data from actual surveys, statistical isotropy is broken due to partial sky coverage and anisotropic noise. This creates a large spurious bispectrum in the squeezed limit due to the correlations between small-scale fluctuations (e.g.\ noise power spectrum) and large-scale effects (e.g.\ scanning pattern of the satellite). As shown in \cite{Jung:2020zne}, the large resulting bias to the integrated bispectrum can be removed using a simple correction term $\ib_\ell^\mathrm{obs} \rightarrow \ib_\ell^\mathrm{obs} - \ib_\ell^\mathrm{lin}$ given by
\begin{equation}
    \label{eq:linear-correction}
    \ib_\ell^\mathrm{lin} = \frac{1}{N_\mathrm{patch}}\sum\limits_{\mathrm{patch}}
        \overbar{\kappa}_\mathrm{patch}^\mathrm{obs} C_{\ell,\mathrm{patch}}^\mathrm{MC}\, .
\end{equation}
This mean-field correction displays a linear dependence on the observed data ($\overbar{\kappa}_\mathrm{patch}^\mathrm{obs}$), while the quadratic term $C_{\ell,\mathrm{patch}}^\mathrm{MC}$ is a Monte-Carlo average of the position-dependent power spectrum from many simulations sharing the same experimental characteristics as the observed data. This is conceptually similar to the standard linear correction of the bispectrum originally introduced in \cite{Creminelli:2005hu}. 

Templates of the expected angular integrated bispectrum for different types of non-Gaussianity can also be computed and fitted to the integrated bispectrum of the data, measured via eq.\ (\ref{eq:ibisp-estimator}). One can show that the exact relation between the full and the integrated bispectrum is given by
\begin{equation}
    \label{eq:ibisp-bisp}
    \begin{split}
        \ib_\ell = 
        & \frac{1}{N_\mathrm{patch}} \frac{1}{4\pi (\fsky^W)^2} \frac{1}{2\ell+1} \sum\limits_{\ell_1 \ell_2 \ell_3} \frac{B_{\ell_1 \ell_2 \ell_3}}{h_{\ell_1 \ell_2 \ell_3}}  \sum\limits_{m_1 m_2 m_3} \begin{pmatrix}
            \ell_1 &  \ell_2 &  \ell_3\\
            m_1 & m_2 & m_3
        \end{pmatrix}\\
        &  \times \sum\limits_{m_4 m_5 m} (-1)^{m} 
        \begin{pmatrix}
            \ell &  \ell_1 &  \ell_4\\
            -m & m_1 & m_4
        \end{pmatrix}
        \begin{pmatrix}
            \ell &  \ell_2 &  \ell_5\\
            m & m_2 & m_5
        \end{pmatrix}
        \sum\limits_{\mathrm{patch}}(w^{\mathrm{patch}}_{\ell_3 m_3})^{*} w_{\ell_4 m_4}^{\mathrm{patch}} w_{\ell_5 m_5}^{\mathrm{patch}}\,,
    \end{split}
\end{equation}
where $\fsky^W$ is the fraction of the sky covered by a patch and the $w_{\ell m}^{\mathrm{patch}}$'s are the harmonic coefficients of a given patch map $W_\mathrm{patch}(\hat{\Omega})$. For the details of this computation and of the remaining equations of this section, we refer the reader to \cite{Jung:2020zne}. 

The sheer number of modes to sum over makes eq.\ \eqref{eq:ibisp-bisp} impossible to calculate in general. However, a careful choice of patches used in the analysis can solve this issue, as first pointed out in \cite{Jung:2020zne}. More specifically, if the patches have a built-in azimuthal symmetry, all the dependence on $m$-multipoles can be integrated out to simplify drastically eq.\ \eqref{eq:ibisp-bisp}. Here, we will focus on the so-called step function patches, a specific type of azimuthally symmetric patches, which are proportional to spherical harmonics and defined by a multipole range $[\lwmin,\lwmax]$
\begin{equation}
    \label{eq:azimuth-patch}
    w_{\ell m}(\hat{\Omega}_0) = \begin{cases} 
    Y_{\ell m}^*(\hat{\Omega}_0) ,~~\lwmin\leq\ell \leq \lwmax \\
    0 ,~~~~~~~~~~~~~  \text{ otherwise.} \
   \end{cases}
\end{equation}
Here, $\hat{\Omega}_0$ defines the position of the patch center. These patches have an important difference with the HEALpix patches that were described earlier. Even if most of their constraining power is located around their centre $\hat{\Omega}_0$, they are actually defined on the full sky.\footnote{Note that $\fsky^W$ cancels out between the observations and the theory. In the different figures presented in this paper showing integrated bispectra, we rescale the integrated bispectra by a $\fsky^W$ defined as the average of the absolute value of a patch map.} This means that the total number of patches and their distribution over the sky is not automatically defined. With our typical choice $\lwmin = 0$ and $\lwmax = 10$ (used to carry out most of the analyses presented in section \ref{sec:simulations}), we found that 192 patches centered in the middle of the pixels of a $\nside=4$ HEALPix map allow us to obtain optimal results (a larger $\lwmax$ would however require more patches). Both types of patches discussed so far (HEALPix and azimuthally symmetric, full-sky patches) are shown in figure \ref{fig:patches}.

Substituting the patch definition (eq.\ \ref{eq:azimuth-patch}) into eq.\ \eqref{eq:ibisp-bisp} the integrated bispectrum becomes
\begin{equation}
    \label{eq:ibisp}
        \ib_\ell 
        = \frac{1}{(4\pi)^3 (\fsky^W)^2} \sum\limits_{\ell_{1,2} = \ell - \lwmax }^{\ell + \lwmax } \sum\limits_{\ell_{3,4,5} = \lwmin } ^{\ell_w}
        B_{\ell_1 \ell_2 \ell_3} \mathcal{F}^{\ell \ell_1 \ell_2}_{\ell_3 \ell_4 \ell_5} \,,
\end{equation}
where $\mathcal{F}^{\ell \ell_1 \ell_2}_{\ell_3 \ell_4 \ell_5}$ depends only on multipole numbers (through Wigner 3$j$ and 6$j$-symbols)
\begin{equation}
    \label{eq:factor-F}
    \begin{split}
        \mathcal{F}^{\ell \ell_1 \ell_2}_{\ell_3 \ell_4 \ell_5} 
         =   &(-1)^{\ell_2 + \ell_4} (2\ell_4+1)(2\ell_5+1)   
        \threej{\ell_1}{\ell_2}{\ell_3}{0}{0}{0}^{-1}
        \threej{\ell}{\ell_1}{\ell_4}{0}{0}{0}
        \threej{\ell}{\ell_2}{\ell_5}{0}{0}{0}
        \threej{\ell_3}{\ell_4}{\ell_5}{0}{0}{0}
        \sixj{\ell_1}{\ell_2}{\ell_3}{\ell_5}{\ell_4}{\ell}.           
    \end{split}
\end{equation}
We recall that by construction the integrated bispectrum is sensitive to the squeezed limit of the bispectrum. In eq.\ \eqref{eq:ibisp}, this is written down explicitly because $\ell_3$, one of the multipole numbers of the bispectrum $B_{\ell_1 \ell_2 \ell_3}$, is in the interval $[\lwmin,\lwmax]$. With our typical choice $\lwmin = 0$ and $\lwmax = 10$, the integrated bispectrum will only probe the multipole configurations where $\ell_3 \leq 10$.\footnote{It is possible to use an $\lwmax$ a few times larger than 10 to probe more configurations, however as pointed out in \cite{Jung:2020zne} this also requires the use of several times more patches in the estimator (eq.\ \ref{eq:ibisp-estimator}) to obtain optimal results in agreement with eq.\ \eqref{eq:ibisp} (for example 768 patches for $\lwmax=20$).}

In most applications, it is also important to compute the covariance matrix of the integrated bispectrum. In the limit of small NG, it was shown in \cite{Jung:2020zne} that, for azimuthally symmetric patches, this can be efficiently derived from the bispectrum covariance, using the formula:
\begin{equation}
    \label{eq:ibisp_covariance}
    \begin{split}
        \mathrm{Covar}(\ib_{\ell^{}}, \ib_{\ell '})
        &=  \frac{1}{(4\pi)^6 (f^W_\mathrm{sky})^4}
        \sum\limits_{\ell_{1,2,3,4,5}^{}} \sum\limits_{\ell'_{1,2,3,4,5}}
        \mathrm{Covar}(B_{\ell_1^{} \ell_2^{} \ell_3^{}}, B_{\ell_1' \ell_2' \ell_3'})
        \mathcal{F}^{\ell^{} \ell_1^{} \ell_2^{}}_{\ell_3^{} \ell_4^{} \ell_5^{}} \mathcal{F}^{\ell' \ell'_1 \ell'_2}_{\ell'_3 \ell'_4 \ell'_5}  \\
       &\equiv  \ic_{\ell\ell'} \,,
    \end{split}
\end{equation}
where the bounds of the summations are the same as in eq.\ \eqref{eq:ibisp}. This formula was used in \cite{Jung:2020zne}, for the analysis of several weak NG signals in the CMB.
In the present work, however, we are interested in the bispectrum of the lensing convergence field and we are no longer in a mild NG regime. Indeed, for mild NG the covariance is almost diagonal (the only non-zero terms are found where $|\ell - \ell'| \leq 2\lwmax$); however, as we also explicitly show in section \ref{sec:covariance}, even very different multipoles $\ell$ and $\ell'$ are correlated in the weak lensing case.

In \cite{Kayo_2012}, a calculation of the weak lensing bispectrum covariance was performed using the Limber approximation; however, here we are interested in the bispectrum squeezed limit, where both $\ell_3^{}$ and $\ell'_3$ are small and this approximation stops to be valid; hence, we cannot directly apply these results to our case. While computing the full weak lensing integrated bispectrum covariance is beyond the scope of the present paper, we discuss this issue further in section \ref{sec:covariance}. There, we offer different possible ways to approach the problem, which we intend to explore in future works.

Meanwhile, as we will see in detail in section \ref{sec:simulations}, we have checked that assuming the weakly non-Gaussian regime already gives a good approximation of the diagonal part of the covariance matrix. In this regime (where $\langle B_{\ell_1^{} \ell_2^{} \ell_3^{}} \rangle \approx 0$), we have the simple expression:
\begin{equation}
    \label{eq:bispectrum-covariance}
    \begin{split}
    \mathrm{Covar}(B_{\ell_1^{} \ell_2^{} \ell_3^{}}, B_{\ell_1' \ell_2' \ell_3'})
    = h_{\ell_1^{} \ell_2^{} \ell_3^{}}^2 C_{\ell_1^{}}C_{\ell_2^{}}C_{\ell_3^{}} ( \delta_{\ell_1^{} \ell_1'}\delta_{\ell_2^{} \ell_2'}\delta_{\ell_3^{} \ell_3'} 
         + \mathrm{permutations} ),
    \end{split}
\end{equation}
depending only on the power spectrum. Substituting this into eq.\ \eqref{eq:ibisp_covariance} gives the integrated bispectrum variance
\begin{equation}
    \label{eq:ibisp_variance}
        \ic_{\ell\ell} \simeq \frac{1}{(4\pi)^6 (\fsky^W)^4} 
                \sum\limits_{\ell_{1,2}^{} = \ell - \lwmax }^{\ell + \lwmax } 
                \sum\limits_{\ell_{3,4,5}^{} = \lwmin } ^{\ell_w}
                \sum\limits_{\ell'_{4,5} = \lwmin } ^{\ell_w}
        h_{\ell_1^{} \ell_2^{} \ell_3^{}}^2  C_{\ell_1^{}}C_{\ell_2^{}}C_{\ell_3^{}} 
        \mathcal{F}^{\ell^{} \ell_1^{} \ell_2^{}}_{\ell_3^{} \ell_4^{} \ell_5^{}}  (\mathcal{F}^{\ell^{} \ell_1^{} \ell_2^{}}_{\ell_3^{} \ell'_4 \ell'_5} + \mathcal{F}^{\ell^{} \ell_2^{} \ell_1^{}}_{\ell_3^{} \ell'_4 \ell'_5} ) \,.
\end{equation}
For simplicity, we omitted some of the permutation terms present in eq.\ \eqref{eq:bispectrum-covariance} which are non-zero only at low $\ell$ (to be exact, when $\ell\leq 2\lwmax$, thus below 20 for the patches discussed above).

\section{Results from all-sky simulations}
\label{sec:simulations}

Accurate simulations of the weak lensing convergence field have been built by Takahashi et al.\ using ray-tracing through N-body simulations (for details see \cite{Takahashi:2017hjr}). Here, we study several sets of these maps at different source redshifts ($z_s=0.5, 1, 1.5, 2$)\footnote{We use the maps with $\nside=4096$ which can be downloaded here at \url{http://cosmo.phys.hirosaki-u.ac.jp/takahasi/allsky_raytracing/nres12.html}. We downgrade their resolution to $\nside=2048$ and impose $\lmax=2000$ for all the analyses presented in this section, unless mentioned otherwise.} both to validate the integrated bispectrum estimator described in section \ref{sec:integrated-bispectrum} for weak lensing studies and to further characterize the squeezed limit of the convergence bispectrum. 

\subsection{Full sky and noiseless case}
\label{sec:ideal-simulations}

\begin{figure}
    \centering
    \includegraphics[width=0.66\linewidth]{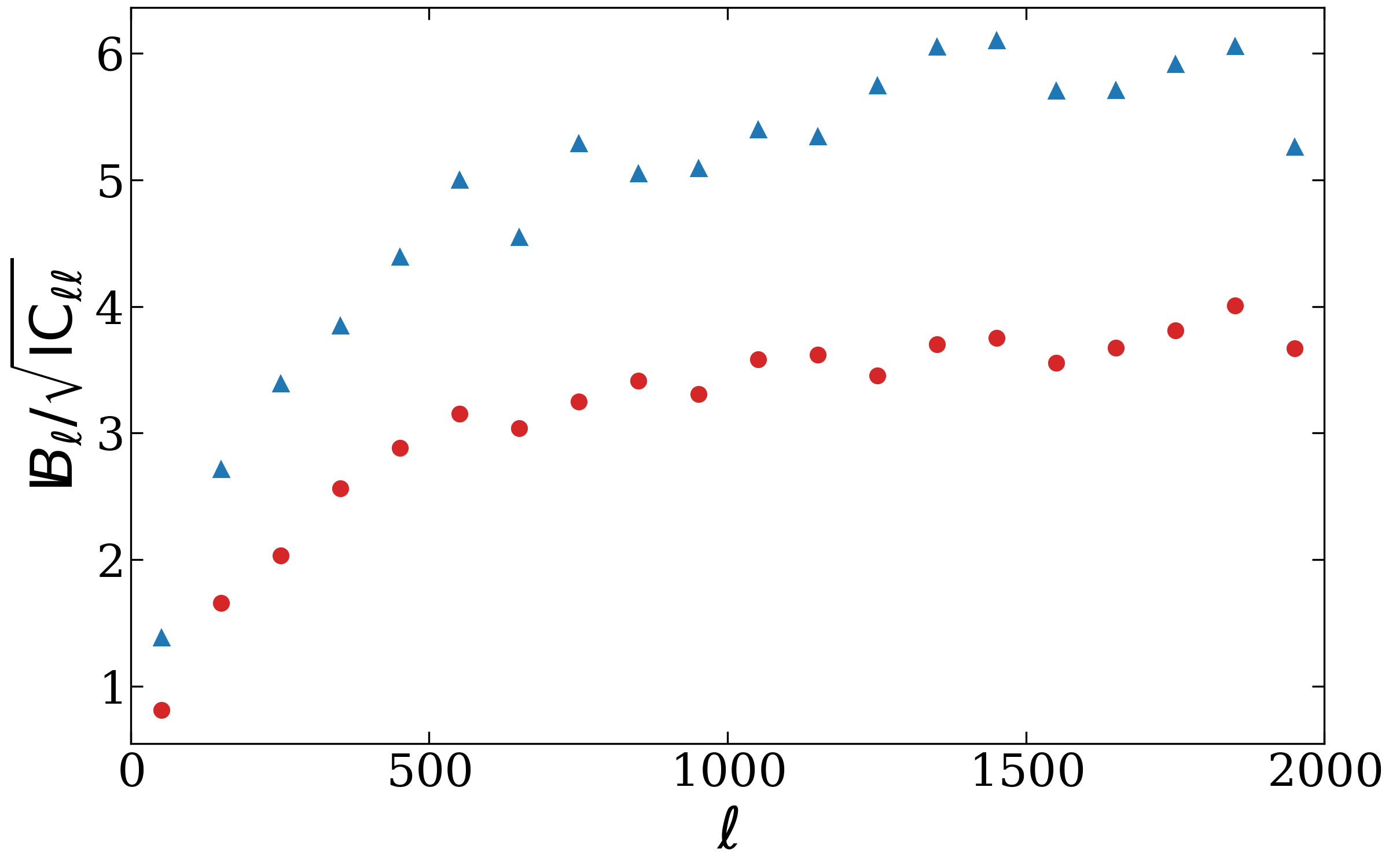}
    \caption{The ratio of the integrated bispectrum to its standard deviation determined from 40 simulations of the weak lensing convergence at a source redshift $z_s=1$ using different types of patches. The blue triangles correspond to HEALPix patches and the red circles to step function patches with $[\lwmin, \lwmax] = [0,10]$. Results are shown using 20 bins.}
    \label{fig:ib_s2n}  
\end{figure}

 The choice of patches and their characteristics (size, shape, number) has a strong effect on the results and constraints obtained with the integrated bispectrum method. Different patches can probe different multipole triplets of the bispectrum with different weights (most of them being squeezed configurations). An illustration of this effect is provided in figure \ref{fig:ib_s2n}, where we show the integrated bispectrum normalized by its standard deviation, extracted from 40 simulations at the source redshift $z_s=1$, using two different types of patches. In both cases, there is a detection of a non-Gaussian signal at several $\sigma$ for $\ell$ above $\sim 100$. The largest signal-to-noise ratio is obtained using the 192 HEALPix \cite{Gorski:2004by} patches described in section \ref{sec:integrated-bispectrum}. These patches correspond to the 192 pixels of a map with $\nside=4$. The other curve is obtained with 192 step function patches, as defined in eq.\ \eqref{eq:azimuth-patch}, with $\lwmin = 0$ and $\lwmax=10$ and with the same centres as the HEALPix patches. With these patches, the integrated bispectrum only probes squeezed configurations of the bispectrum with $\ell_3 \leq 10$ (see eq.\ \ref{eq:ibisp}). The step function patches give a smaller signal-to-noise ratio because the HEALPix pixels are sensitive to more modes (other squeezed configurations with a larger $\ell_3$). However, both the integrated bispectrum and its variance can be computed analytically in the azimuthally symmetric case, making the step function patches more adapted for a fast and exact characterization of the squeezed limit of the bispectrum, with a possible comparison to theoretical predictions.

\begin{figure}
    \centering
    \includegraphics[width=0.99\linewidth]{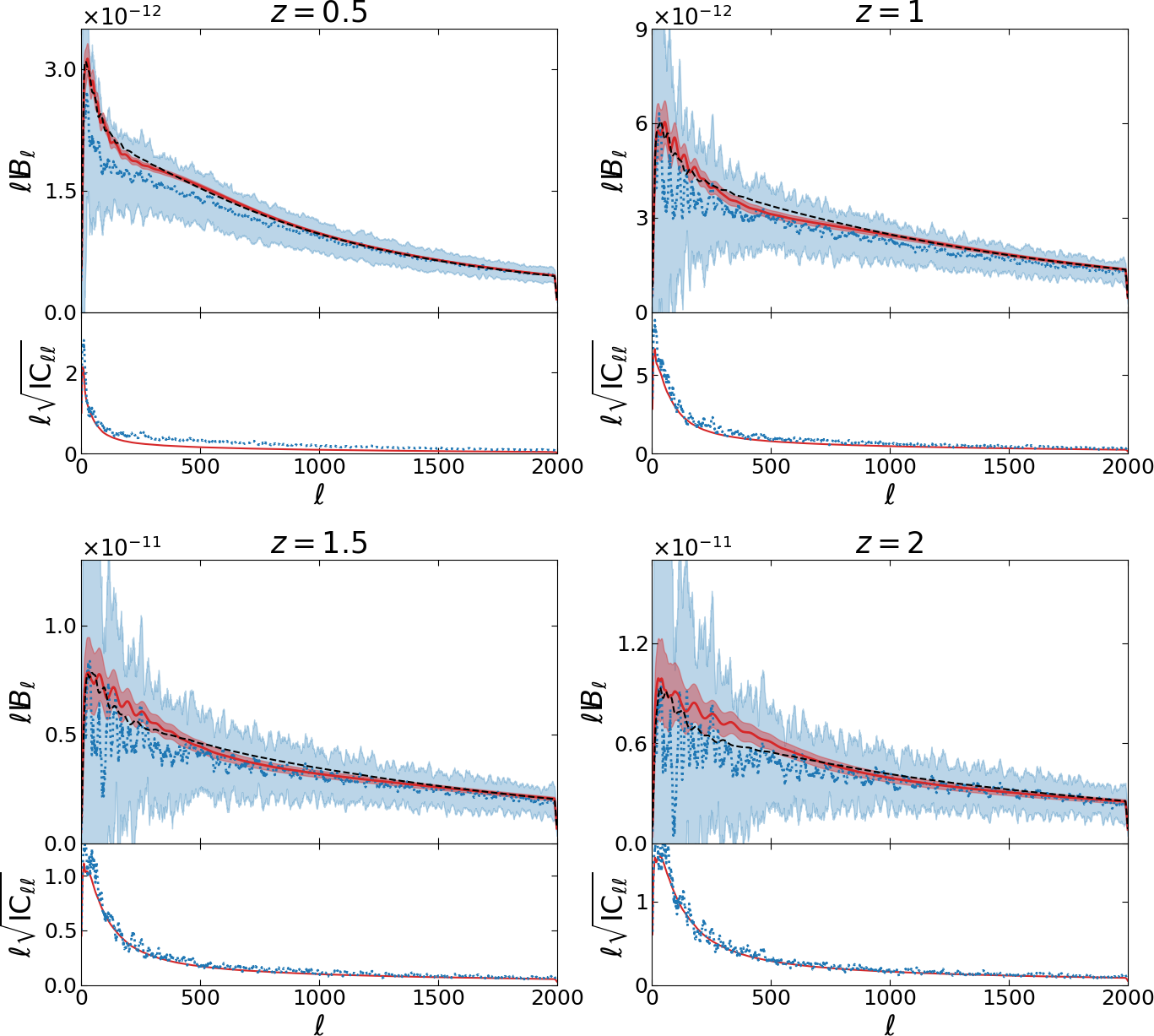}
    \caption{The integrated bispectrum of the weak lensing convergence at the source redshifts ${z_s=0.5, 1, 1.5, 2}$ using step function patches ($[\lwmin, \lwmax] = [0,10]$). In blue (dotted lines), the averaged integrated bispectra from sets of 40 simulations, determined using the estimator eq.\ \eqref{eq:ibisp-estimator}, with their error bars (standard deviations also determined from these simulations). In red (solid lines), the theoretical templates computed using eqs.\ \eqref{eq:ibisp} and \eqref{eq:ibisp_covariance} with the standard error on the mean expected from 40 maps (attached to the theory curve for clarity). The black dashed lines are also theoretical integrated bispectra, where post-Born corrections are included. The bottom panel of each plot compares the measured and theoretical standard deviations. All quantities have been multiplied by a factor $\ell$ for readability.}
    \label{fig:ib_4z_lw10}  
\end{figure}

In figure \ref{fig:ib_4z_lw10}, we show the integrated bispectra and their corresponding error bars at the source redshifts $z_s=0.5, 1, 1.5, 2$ measured from sets of 40 simulations (in blue) and their corresponding theoretical predictions using eqs.\ \eqref{eq:ibisp} and \eqref{eq:ibisp_covariance} (in red). Note that the red error bars of each top panel correspond to the expected standard error on the mean from 40 maps. Measured and theoretical standard deviations are compared in each bottom panel. The theoretical power spectra used to evaluate the standard deviations $\sqrt{\ic_{\ell\ell}}$ are the averaged power spectra of these sets of simulations. The theoretical bispectra, necessary to obtain the integrated bispectrum templates, are computed using eq.\ \eqref{eq:convergence-bispectrum}. The integrated bispectrum templates shown here are fully described in section \ref{sec:weak-lensing}. For each source redshift, there is a clear detection at several $\sigma$ of a non-Gaussian signal. There is a good agreement between the observed results and the theoretical predictions, with however a small discrepancy of a few percent. The theoretical templates are always slightly larger than the integrated bispectra from simulations. Concerning the standard deviations, there is only a large difference (a factor $\sim 2$) in the most non-Gaussian case ($z_s=0.5$) where the assumption of weak non-Gaussianity necessary to derive eq.\ \eqref{eq:bispectrum-covariance} is clearly broken. Otherwise, they are of the same order (on average 10\% larger from the simulations, which can also be explained by the fact that these maps are also not Gaussian). In figure \ref{fig:ib_4z_lw10}, we also show the integrated bispectrum templates including the post-Born correction. At these redshifts, its effect is several times smaller than the standard deviation. However, when studying the average of many maps, the difference is not negligible (it is actually of the same order as the standard error on the mean of 40 maps).

\begin{figure}
    \centering
    \includegraphics[width=0.66\linewidth]{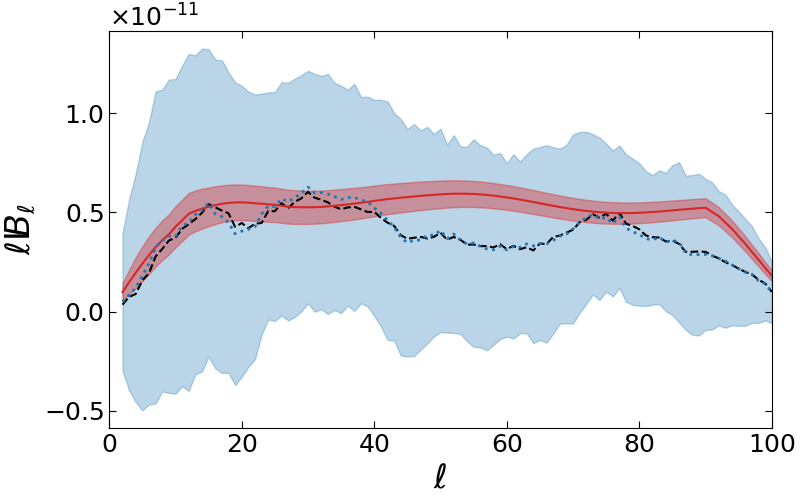}
    \caption{The integrated bispectrum of the weak lensing convergence at the source redshift $z_s=1$. The blue dotted lines correspond to the measured averaged integrated bispectrum from 40 simulations using the estimator eq.\ \eqref{eq:ibisp-estimator}, and its corresponding error bars while the red solid lines are the theoretical predictions computed using eqs.\ \eqref{eq:ibisp} and \eqref{eq:ibisp_covariance} (the red error bar being the standard error on the mean). The black dashed line is also computed using eq.\ \eqref{eq:ibisp}, but the substituted bispectrum template is the measured bispectrum of the simulations (using the estimator eq.\ \ref{eq:binned-bispectrum}, introduced later).} 
    \label{fig:ib_full_bispectrum}  
\end{figure}

It is important to verify that the small, but non-negligible, discrepancies obtained with the integrated bispectrum estimator (eq.\ \ref{sec:integrated-bispectrum}) are actually not an issue of the estimator itself. This requires checking that the estimator gives compatible results with the expected integrated bispectrum shape (eq.\ \ref{eq:ibisp}) when the exact, full bispectrum in the observational data is known. To this end, we measure the full bispectrum $B_{\ell_1 \ell_2 \ell_3}^\mathrm{obs}$ of the maps at $z_s=1$ (using the estimator eq.\ \ref{eq:binned-bispectrum} described in the next section) after downgrading the resolution of the maps to $\nside=128$ with an $\lmax=100$ to keep the required computational time reasonable. Then we calculate the expected integrated bispectrum shape by using this $B_{\ell_1 \ell_2 \ell_3}^\mathrm{obs}$ as a theoretical template in eq.\ \eqref{eq:ibisp}. Finally one has just to compare this result to the integrated bispectrum directly determined by applying the standard estimator to the simulations (power spectra in patches, eq.\ \ref{eq:ibisp-estimator}). In figure \ref{fig:ib_full_bispectrum}, we show the outputs of the two methods. There is very good agreement between the two integrated bispectra computed as described above (integrated bispectrum estimator applied to simulations vs.\ full bispectrum estimator, followed by the theoretical computation of the corresponding integrated bispectrum), whereas we do not have perfect agreement of either of these measurements with the pure theoretical template (using the bispectrum eq.\ \ref{eq:convergence-bispectrum}). This confirms that the discrepancies between the theoretical template and the measured results, highlighted in figure \ref{fig:ib_4z_lw10}, are not related to issues in the integrated bispectrum pipeline. Note that the similar discrepancy has been also reported by multiple works such as \cite{Namikawa:2018bju}. As shown in \cite{Takahashi:2019hth}, the lensing bispectrum in the squeezed configuration has a discrepancy between the fitting formula and simulations by up to roughly 10$\%$. Although this discrepancy appears irrespective of the box size and resolution of the simulation, there are still several possibilities to explain this discrepancy such as the lens-shell thickness as similar to the power spectrum and the Limber approximation of the bispectrum calculation.

\subsection{A more realistic case}

\begin{figure}
    \centering
    \includegraphics[width=0.99\linewidth]{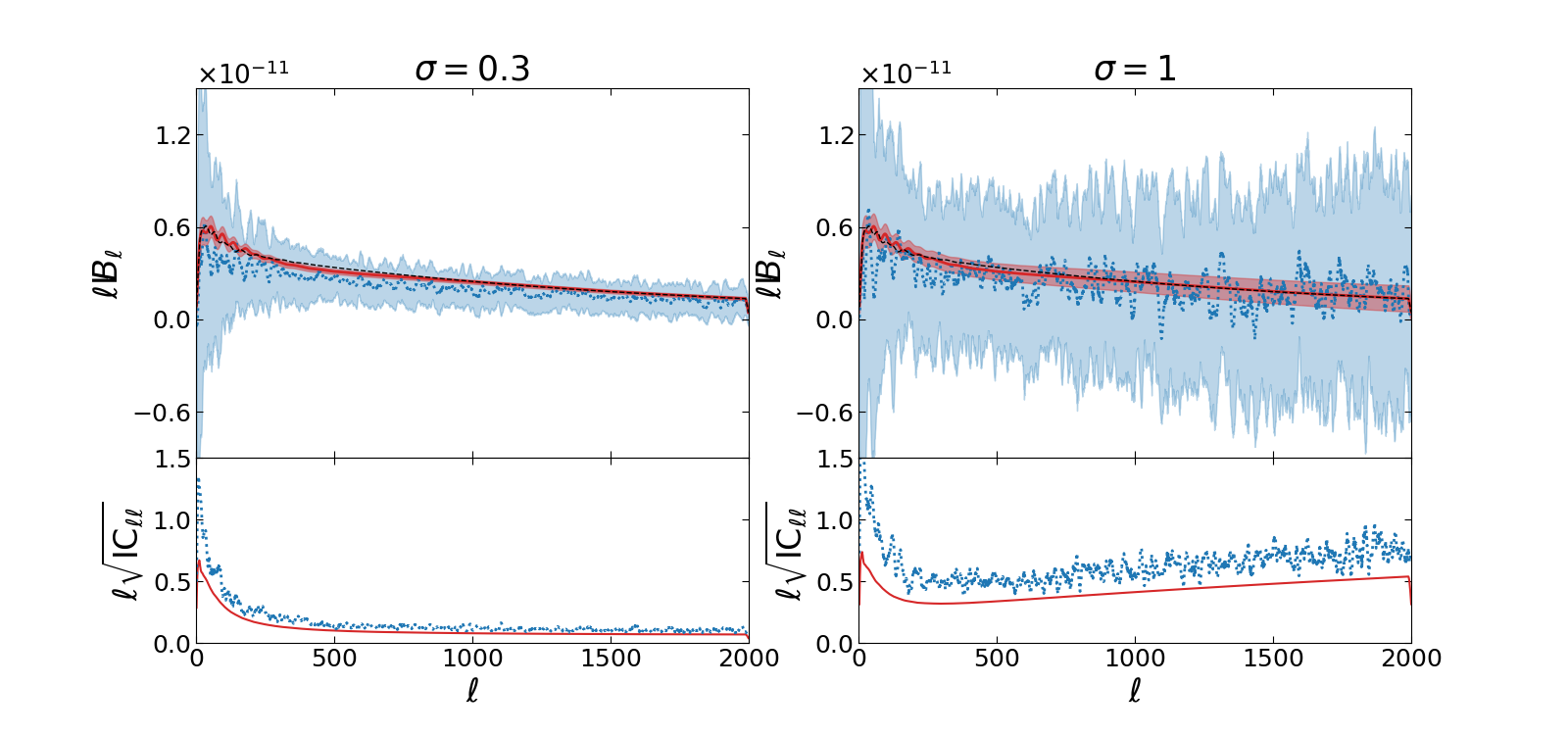}
    \caption{Similar to figure \ref{fig:ib_4z_lw10}, at the source redshift $z_s=1$ for different levels of Gaussian noise as defined in eq.\ \eqref{eq:nl} and using a pseudo seudo \euclid\ mask ($\fsky=0.35$). The observed integrated bispectra have been multiplied by $1/\fsky$. In the bottom panels, the theoretical standard deviations (solid red lines) is given in the full sky case.}
    \label{fig:ib_zs16}  
\end{figure}

While the previous results are obtained in the ideal case (full-sky maps without noise), the integrated bispectrum method can also be applied to more realistic observations. To illustrate this, we use the same 40 simulations at $z_s=1$ as before, to which we add some \euclid-like characteristics. First, we use the same pseudo \euclid\ mask as in \cite{Munshi:2020tzm} which hides both the galactic and elliptic planes ($\fsky=0.35$). Then, we assume Gaussian noise, with a noise power spectrum amplitude given by:
\begin{equation}
    \label{eq:nl}
    n_\ell = \frac{\sigma^2}{\bar{n}},
\end{equation}
where  the galaxy number density $\bar{n}$ (typically $\bar{n}=30$ arcmin$^{-2}$ for Euclid \cite{Laureijs2011}) should be expressed in inverse steradians.  If galaxy clustering is used to estimate the convergence, then the shot noise is $1/\bar{n}$ (i.e.  $\sigma=1$), whereas for size and flux, the error typically corresponds to $\sigma=0.8$ \citep{Alsing2015}. To span the range, we present results for $\sigma = 0.3$ and $\sigma=1$, where the lower value applies to shape noise for shear measurements \citep{Hildebrandt2017}, although our calculations are not directly applicable to the shear case, so this is illustrative.  In figure \ref{fig:ib_zs16}, we show the results obtained after masking the simulations ($\fsky=0.35$) and adding Gaussian noise realizations. Our pipeline includes a diffuse inpainting procedure (standard in CMB non-Gaussianity estimation like the Planck 2018 analysis \cite{Akrami:2019izv}) to smooth the edges of the mask and avoid power leaking between different multipoles. Masking also breaks isotropy and requires using the correction term defined in eq.\ \eqref{eq:linear-correction} to keep results as close to optimality as possible (the relatively small increase of variance shown in figure \ref{fig:ib_zs16} is entirely due to the fact that there is less sky to observe and is unrelated to the loss of isotropy). In the low noise case, there is a detection of the non-Gaussian signal at more than 1-$\sigma$ at every multipole. In the other plot we can see that the expected signal-to-noise ratio becomes much lower, making it is necessary to integrate over $\ell$-modes in order to achieve a large enough signal-to-noise level to confirm the presence of non-Gaussianity and characterize it. However, this requires the knowledge of the full covariance matrix, which adds a level of difficulty to this analysis, see section \ref{sec:covariance} for more details.

\subsection{CMB lensing}
\label{sec:cmb-lensing}

\begin{figure}
    \centering
    \includegraphics[width=0.66\linewidth]{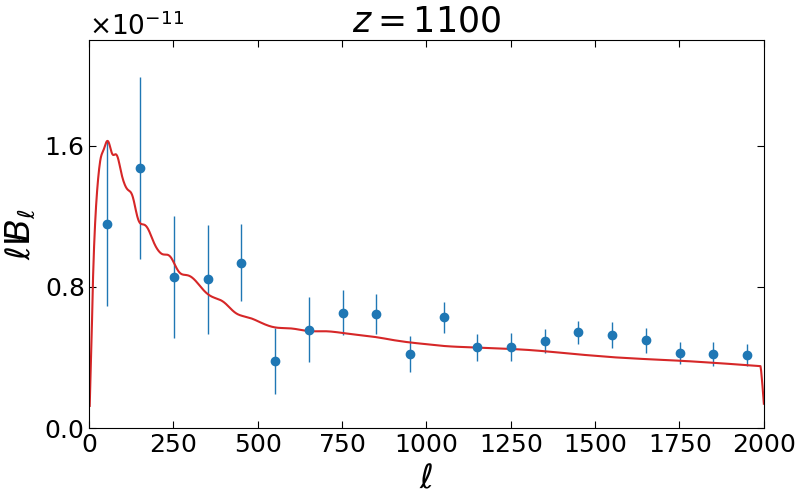}
    \caption{The integrated bispectrum of weak lensing convergence at ${z_s=1100}$ using step function patches ($[\lwmin, \lwmax] = [0,10]$). The red solid line is the theoretical expectation. The blue circles correspond to the averaged integrated bispectrum from 108 simulations, using 20 bins in multipole space to represent it. The vertical blue lines are the standard errors of the mean of each bin determined from the simulations. All quantities have been multiplied by a factor $\ell$ for readability.}
    \label{fig:ib_z1100}  
\end{figure}

Takahashi et al.\ also provided lensing maps at the time of the emission of the CMB \cite{Takahashi:2017hjr}. We apply the same integrated bispectrum method as before with the step function patches to the full set of 108 maps. Results are shown in figure \ref{fig:ib_z1100}. There is a good agreement between the theory, which by default includes the post-Born correction here, and the measured results. As expected, the non-Gaussian signal is much smaller than at the low redshifts studied in section~\ref{sec:simulations}. 

\section{The squeezed limit of the binned bispectrum}
\label{sec:binned-bispectrum}

Another standard approach to study weak lensing non-Gaussianity is the binned bispectrum estimator. It has been used recently in \cite{Munshi:2019csw,Takahashi:2019hth} on the same all-sky simulations as those studied in section \ref{sec:simulations}. Here, we want to apply the binned bispectrum approach as an additional check of the integrated bispectrum results. Hence the focus is on the squeezed limit of the bispectrum.

\subsection{The binned bispectrum estimator}

The binned bispectrum method is based on the simple idea that if a bispectrum template is relatively smooth it is possible to bin it in multipole space with a negligible loss of information. The number of binned bispectrum modes scales as $N_\mathrm{bin}^3$ (where $N_\mathrm{bin}$ is the number of bins) instead of $\lmax^3$ for a full bispectrum. This represents a huge decrease in the number of configurations to consider, when using a typical $N_\mathrm{bin}\sim 10-100$, compared to a standard $\lmax \sim 2000$ (or even more) of recent experiments. Defining the binning of multipole space as $\Delta_i = [\ell_i, \ell_{i+1} -1]$ where $i$ goes from 0 to $N_\mathrm{bin}$, $\ell_0 = \lmin$ and $\ell_{N_\mathrm{bin}} = \lmax + 1$, the binned bispectrum estimator can be written as
\begin{equation}
    \label{eq:binned-bispectrum}
    B_{i_1 i_2 i_3}^\mathrm{obs} = \frac{1}{N_{i_1 i_2 i_3}} \sum\limits_p^{N_\mathrm{pix}}
     M_{i_1}^\mathrm{obs}(\hat\Omega_p) M_{i_2}^\mathrm{obs}(\hat\Omega_p) M_{i_3}^\mathrm{obs}(\hat\Omega_p)\,,
    \quad\text{with}~~
    M_i(\hat\Omega_p)=\sum\limits_{\ell\in\Delta_i}\sum\limits_{m=-\ell}^\ell \kappa_{\ell m} Y_{\ell m}(\hat\Omega_p)\,,
\end{equation}
where $N_{i_1 i_2 i_3}$ is a normalization factor depending on whether a reduced or an angle-averaged bispectrum is needed. 

\subsection{Results with the binned bispectrum}

\begin{figure}
    \centering
    \includegraphics[width=0.49\linewidth]{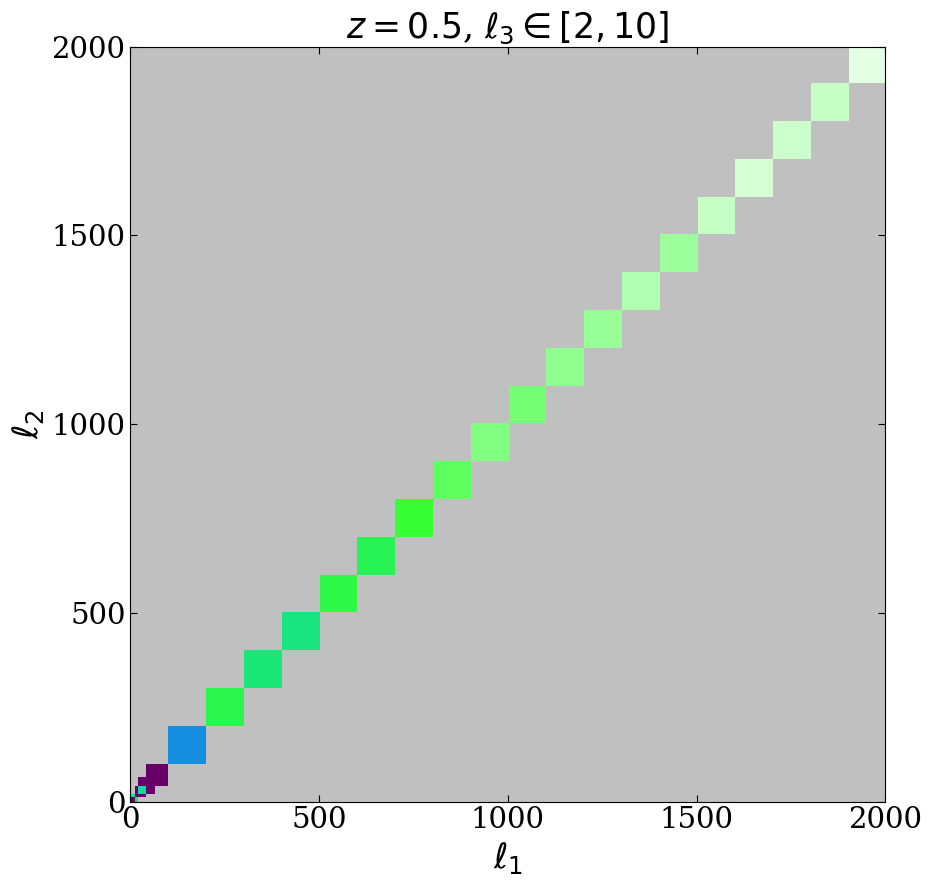}
    \includegraphics[width=0.49\linewidth]{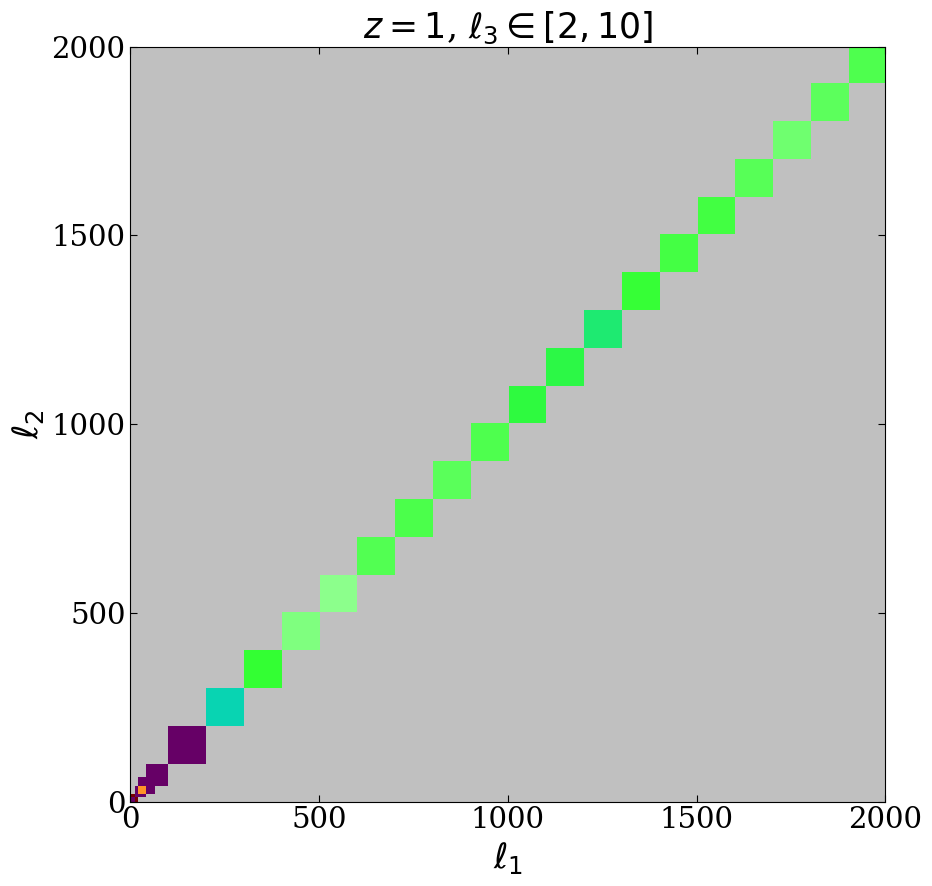}
    \includegraphics[width=0.49\linewidth]{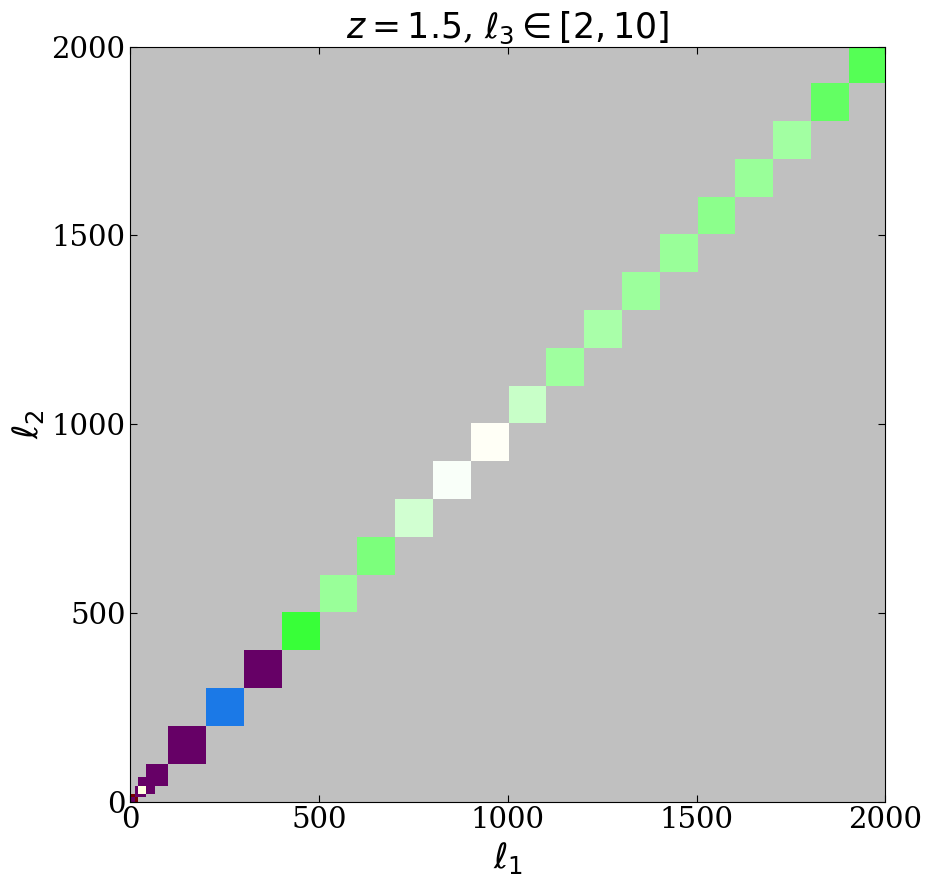}
    \includegraphics[width=0.49\linewidth]{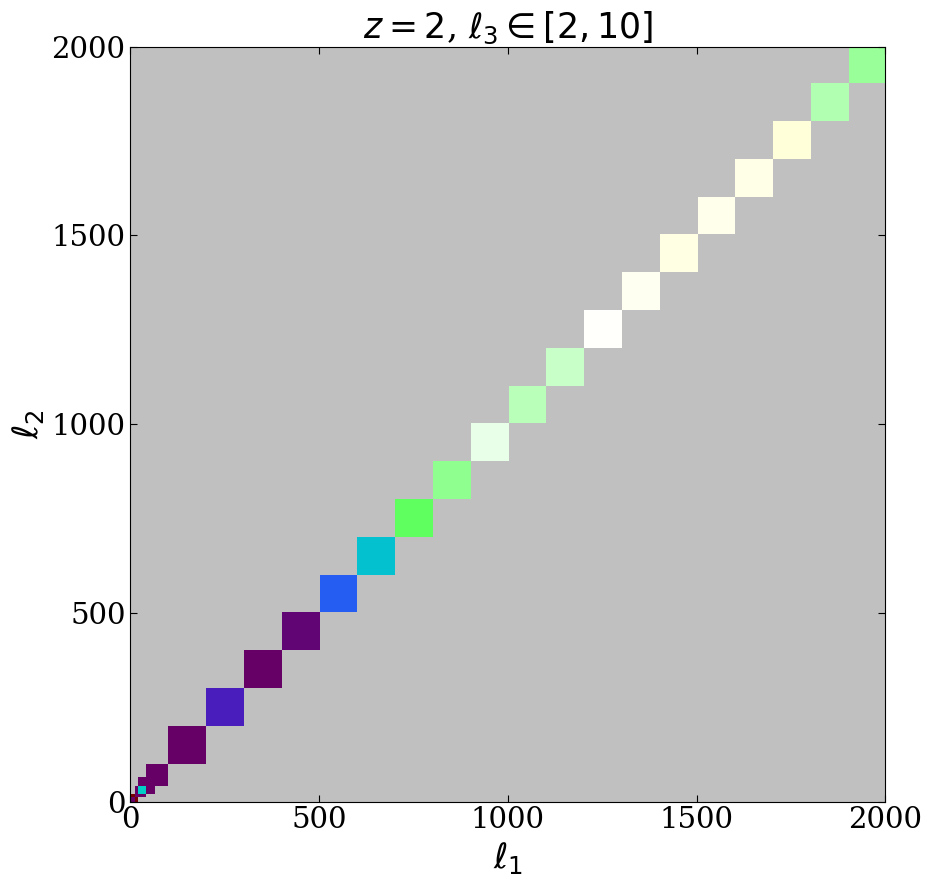}
    \includegraphics[width=0.5\linewidth]{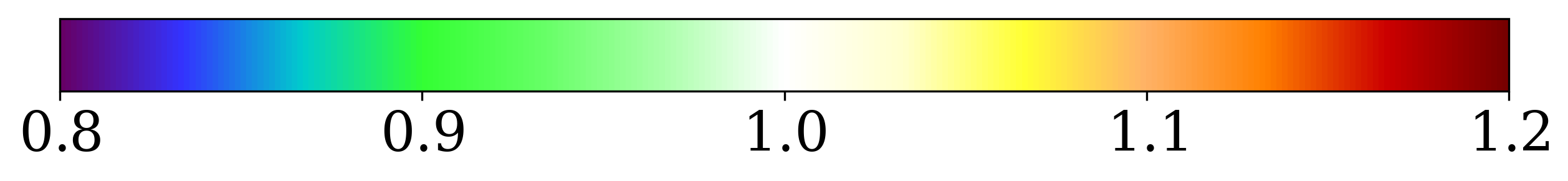}
    \caption{The binned bispectrum of weak lensing convergence at the source redshifts $z_s=0.5, 1, 1.5, 2$. The measured binned bispectra are determined from the same sets of 40 simulations as figure \ref{fig:ib_4z_lw10} and divided by their theoretical counterparts. Only the squeezed limit ($\ell_3 \leq 10$) is shown, other configurations can be seen in figures \ref{fig:binbisp_zs1} and \ref{fig:binbisp}. A total of 24 bins, delimited by $[2,11,21,41,66,100,200,300,400,\dots,1900,2000]$ have been used.} 
    \label{fig:binbisp_4z}  
\end{figure}

\begin{figure}
    \centering
    \includegraphics[width=0.49\linewidth]{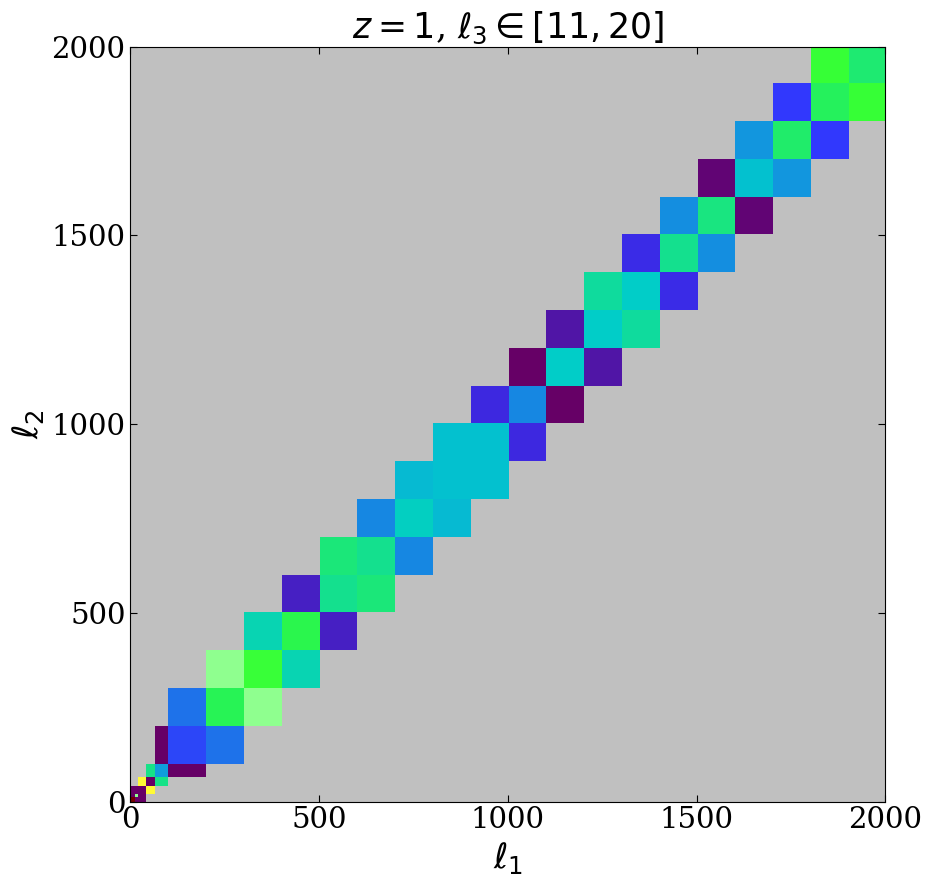}    
    \includegraphics[width=0.49\linewidth]{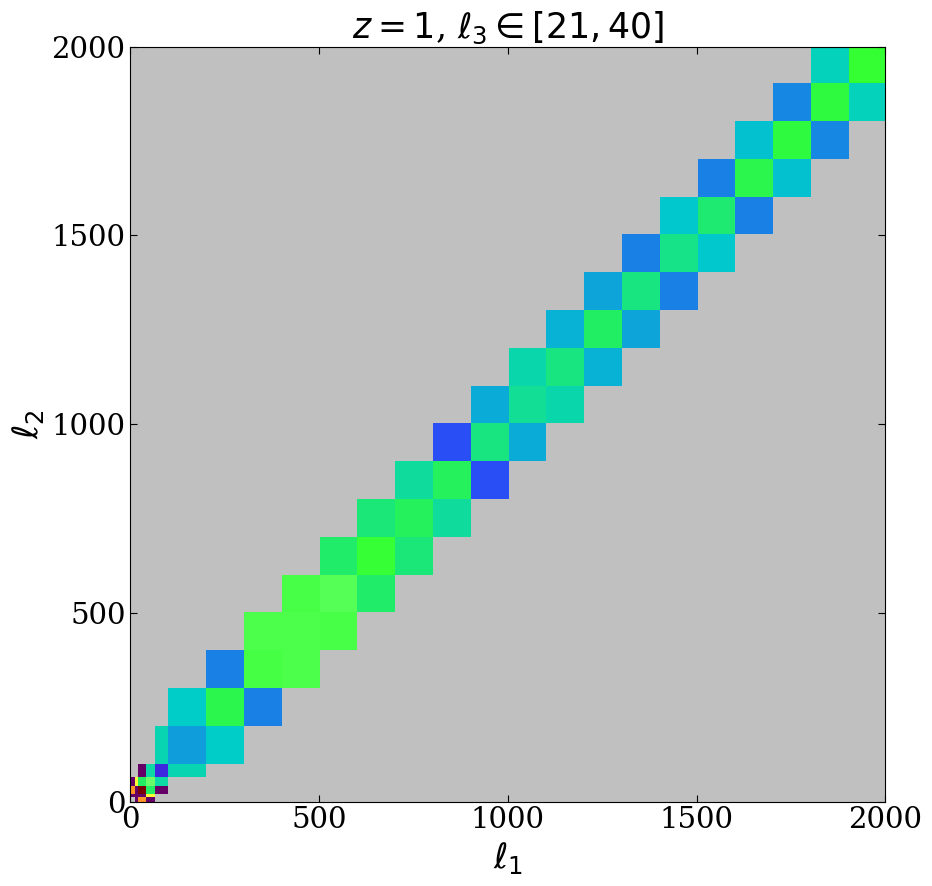}
    \includegraphics[width=0.49\linewidth]{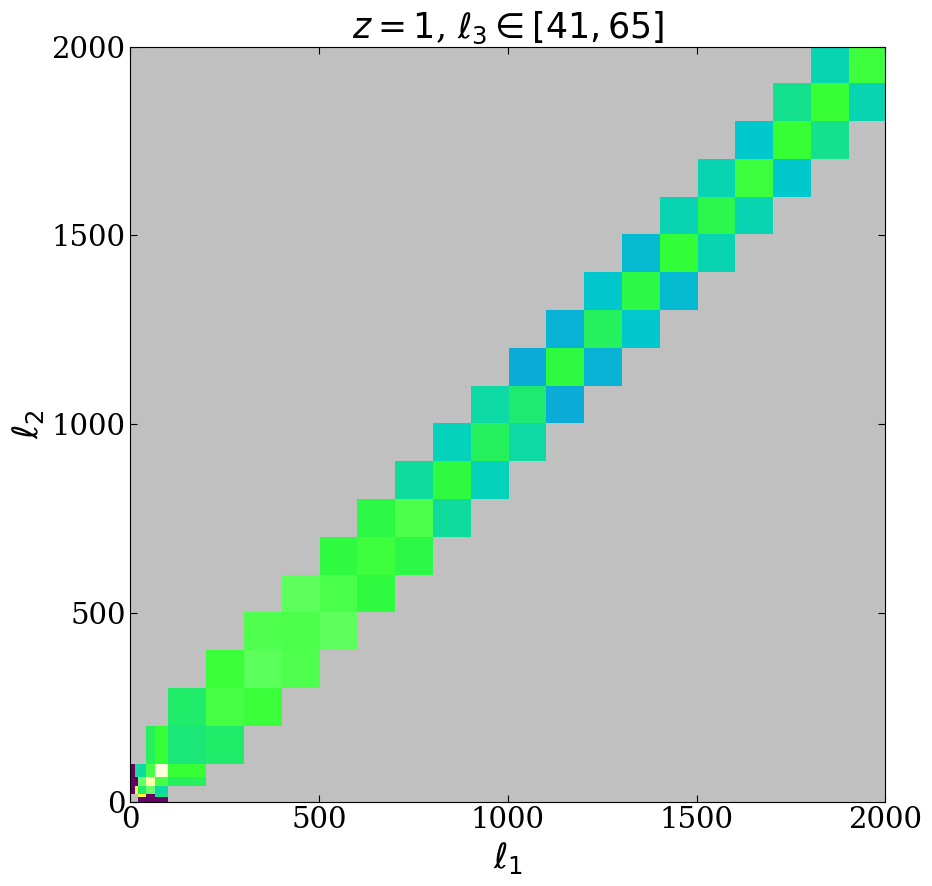}
    \includegraphics[width=0.49\linewidth]{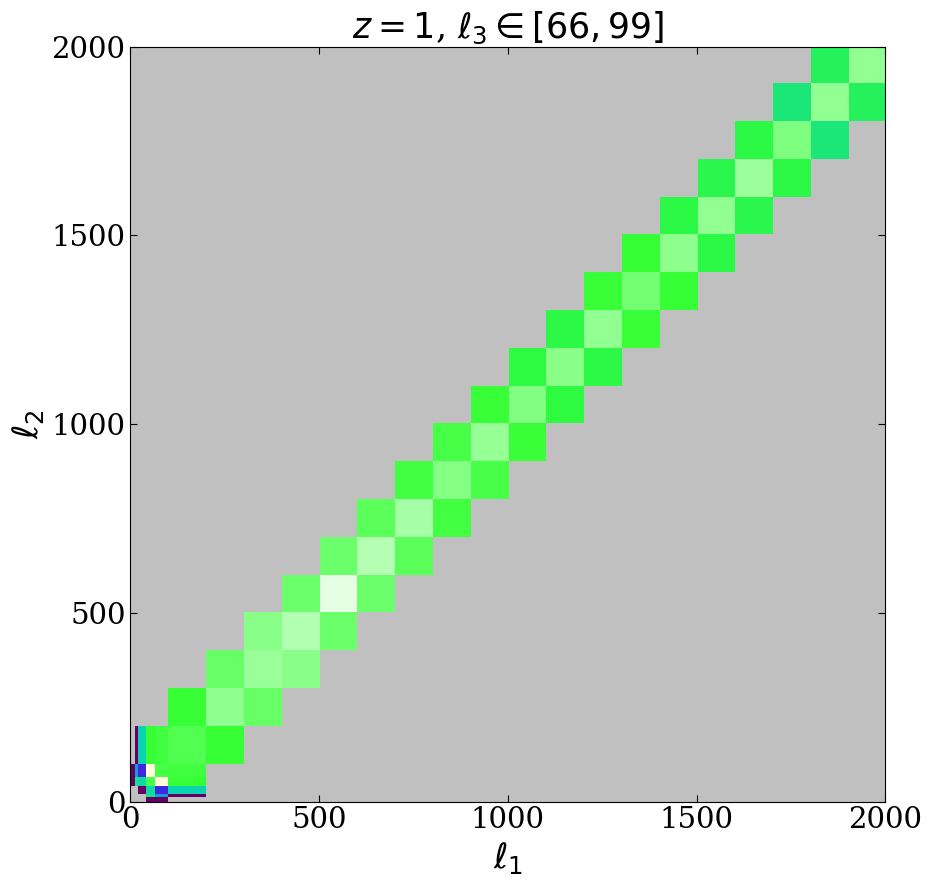}
    \includegraphics[width=0.5\linewidth]{figures/binned_bispectrum/colorbar_bispectrum_ratio.png}
    \caption{Similar to figure \ref{fig:binbisp_4z} for other squeezed modes ($\ell_3$ between 11 and 99) at the source redshift $z_s=1$.}
    \label{fig:binbisp_zs1}  
\end{figure}

We apply this estimator to the same sets of maps as in section \ref{sec:simulations}. The squeezed limit of the obtained binned bispectra is shown in figures \ref{fig:binbisp_4z} and \ref{fig:binbisp_zs1}. To be exact, we plot the ratio of the averaged binned bispectrum from simulations to the binned theoretical template at the source redshifts $z_s=0.5, 1, 1.5, 2$. On the first figure, $\ell_3$ is fixed in the interval $[2,10]$, which corresponds to the same $\ell_3$-range probed by the step function patches used in section \ref{sec:simulations}. This allows for a comparison of the binned bispectra with the integrated bispectra shown earlier in figure \ref{fig:ib_4z_lw10}. As expected, there is a similar discrepancy between measured and theoretical bispectra (theoretical templates are roughly $10\%$ larger in most of the squeezed configurations). The second figure, where $\ell_3$ is fixed to larger but still small values (thus still squeezed modes), confirms this trend. Note however that when $\ell_3$ increases, the discrepancy becomes smaller meaning that the template (eq.\ \ref{eq:convergence-bispectrum}) becomes an accurate description of the non-Gaussian signal outside of the squeezed limit. This can also be seen in appendix \ref{ap:binned-bispectrum-configurations} where we show more configurations with a larger $\ell_3$. The error bars on these binned bispectra, which are not shown in figures \ref{fig:binbisp_4z} and \ref{fig:binbisp_zs1}, are small when considering the average of 40 simulations. However, it becomes important to predict and characterize these error bars for the analysis of one observational data map. Working with large bins as here (width of 100 for $\ell\geq100$) requires more than the bispectrum variance eq.\ \eqref{eq:bispectrum-covariance}, even as a first approximation, because of the non-negligible correlations between different modes, especially in the squeezed limit. This issue is discussed in section \ref{sec:covariance}. 

\section{The next step: covariance matrix}
\label{sec:covariance}

In section \ref{sec:simulations}, we have shown that the integrated bispectrum variance $\ic_{\ell\ell}$ is well described by eq.\ \eqref{eq:ibisp_variance}, where we assume to be in the regime of weak non-Gaussianity. However, we verify that this approximation no longer works well enough when we look at off-diagonal covariance terms. Weak lensing convergence maps are indeed significantly non-Gaussian. An accurate description of the bispectrum and integrated bispectrum covariances thus requires including higher-order connected correlation functions in the calculation. A calculation of this type was performed in \cite{Kayo_2012}, using however the Limber approximation. The same approach is not sufficient here because we are mainly interested in the squeezed limit, where this approximation is not valid. While the full computation of the squeezed bispectrum covariance is beyond the scope of this paper, we can already study at this level the level of correlation between different configurations (i.e., off-diagonal covariance terms) in the different analyses performed in section \ref{sec:simulations} and \ref{sec:binned-bispectrum}.

\begin{figure}
    \centering
    \includegraphics[width=0.49\linewidth]{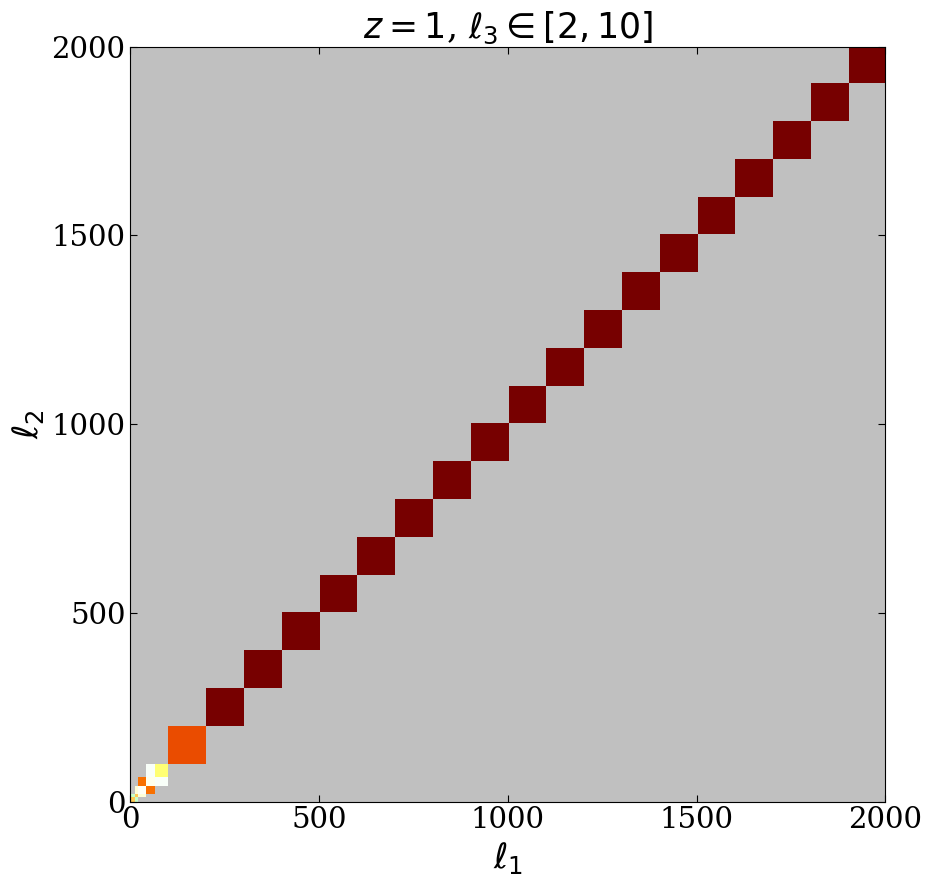}
    \includegraphics[width=0.49\linewidth]{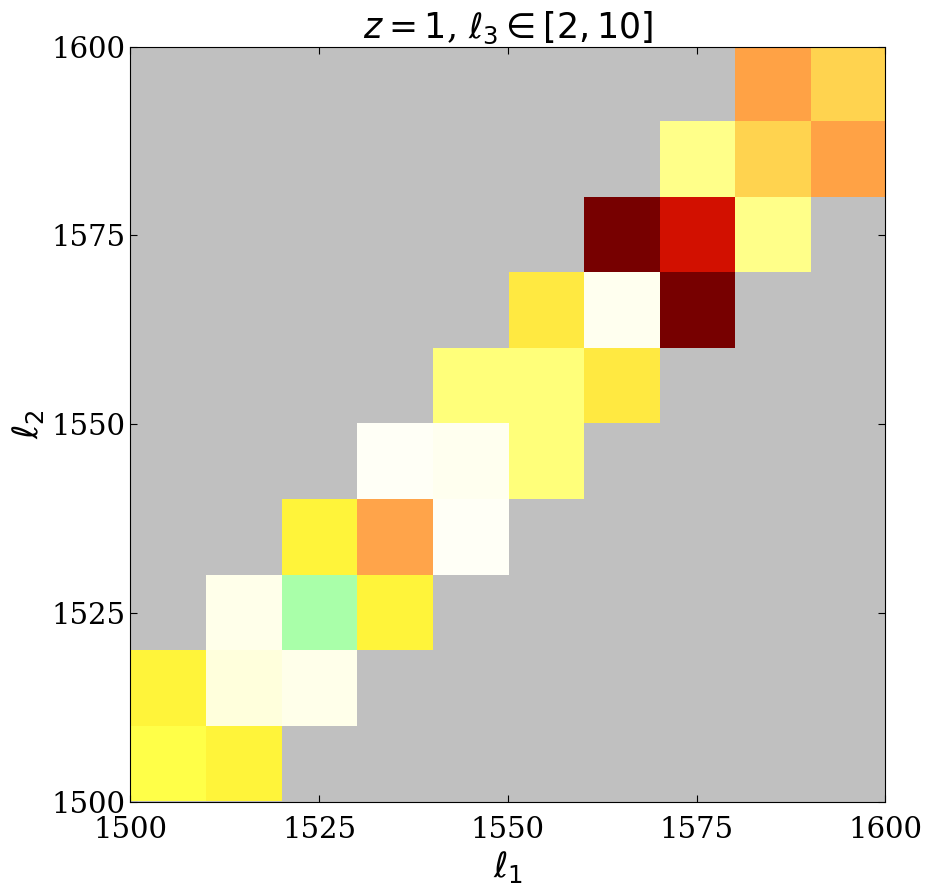}
    \includegraphics[width=0.4\linewidth]{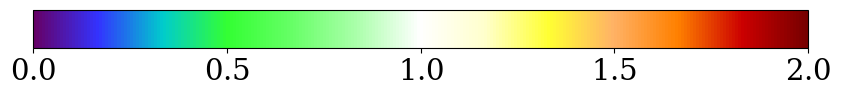}
    \caption{The variance of the binned bispectrum estimator from a set of 40 simulations at the source redshift $z_s=1$, divided by its theoretical prediction assuming that eq.\ \eqref{eq:bispectrum-covariance} holds (weak non-Gaussianity regime). On the left panel, the same binning as in figures \ref{fig:binbisp_4z}, \ref{fig:binbisp_zs1} and \ref{fig:binbisp} has been used while on the right panels only a smaller part of multipole space is shown ($\ell_1, \ell_2 \in [1500,1600]$ with smaller bins defined by $[1500,1510,1520,\dots,1590,1600]$.)}
    \label{fig:ratio-variance}  
\end{figure}

A simple example is the covariance matrix of the binned bispectrum estimator used in section \ref{sec:binned-bispectrum}. Because of the large bins (width of 100 for $\ell > 100$), each triplet of bins contain many different correlated configurations. As can be seen in figure \ref{fig:ratio-variance} (left panel), the estimated covariance matrix from 40 convergence simulations at $z_s=1$ is, in the squeezed limit, more than two times larger that its theoretical expectation computed in the diagonal case. On the right panel, using smaller bins (width of 10), the difference becomes much smaller for most of the configurations. This shows that, taken individually, the effect of non-diagonal terms is rather small; however, when integrating over many modes (e.g.\ using large bins), their summed contribution quickly becomes the dominant one.

\begin{figure}
    \centering
    \includegraphics[width=0.66\linewidth]{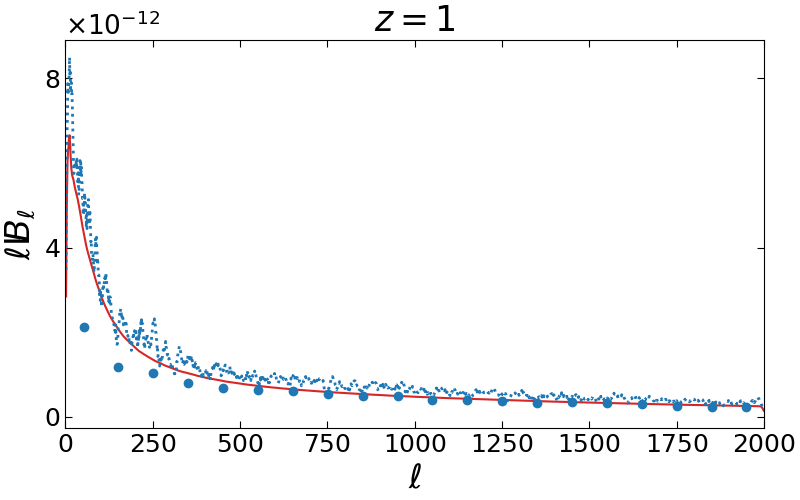}
    \caption{The integrated bispectrum variance $\ic_{\ell\ell}$ of the weak lensing convergence at $z_s=1$ using step function patches. The solid red line corresponds to the theoretical prediction using eq.\ \eqref{eq:ibisp_variance}. The blue dotted line is the variance estimated from 40 simulations, while the blue circles are also determined from these simulations after compressing the integrated bispectrum information to 20 bins.}
    \label{fig:binned-ibisp}  
\end{figure}

\begin{figure}
    \centering
    \includegraphics[width=0.49\linewidth]{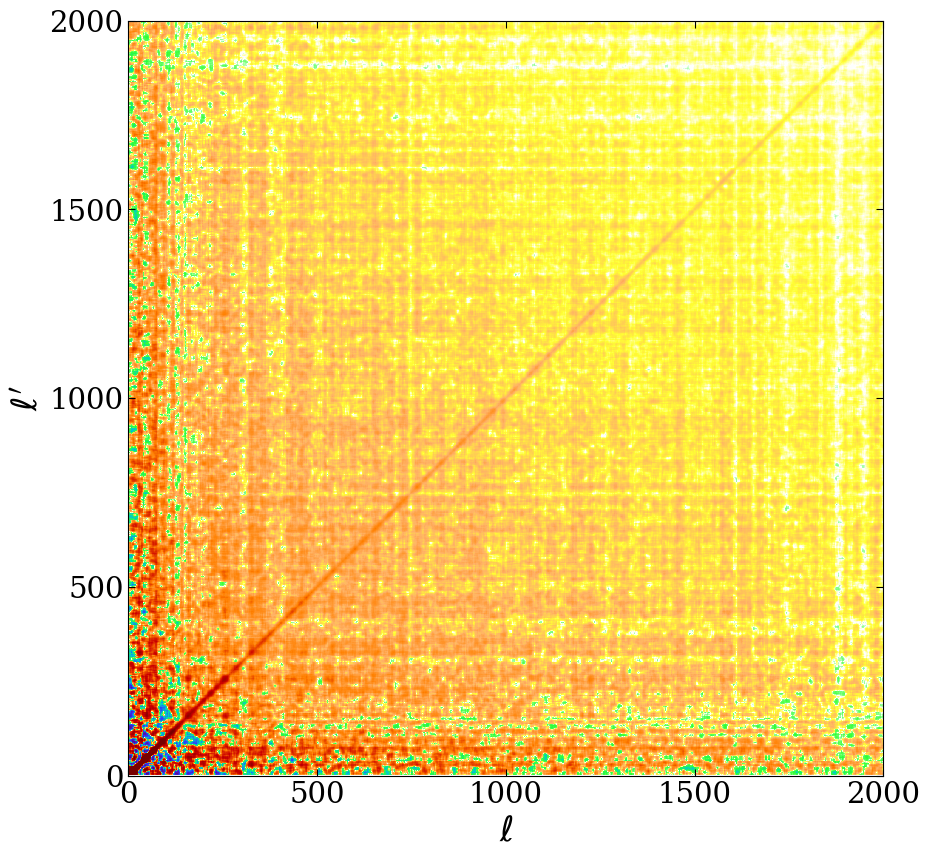}
    \includegraphics[width=0.49\linewidth]{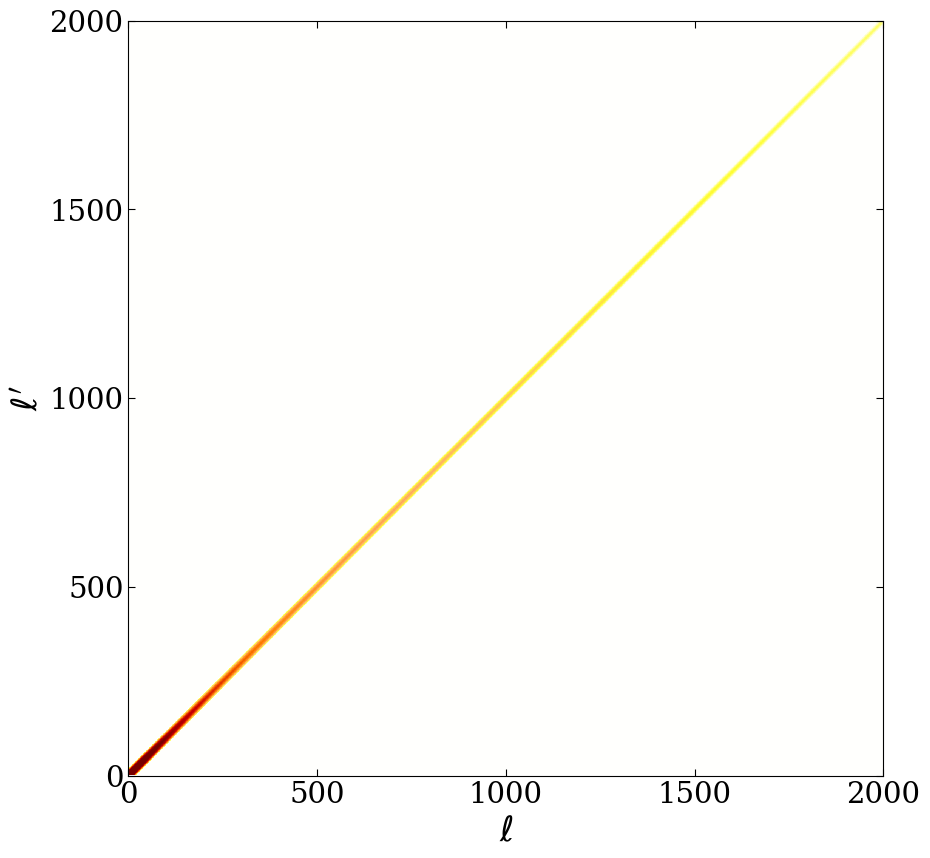}
    \includegraphics[width=0.4\linewidth]{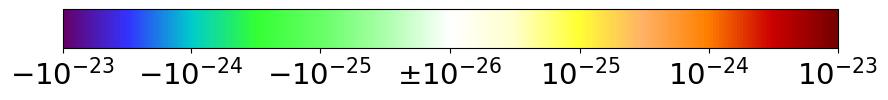}
    \caption{The integrated bispectrum covariance $\ic_{\ell\ell'}$ of the weak lensing convergence at $z_s=1$ using step function patches. On the left, the covariance is estimated from the 40 simulations studied in section \ref{sec:simulations}. On the right, the theoretical covariance computed using eq.\ \eqref{eq:ibisp_covariance} if the bispectrum covariance is given by eq.\ \eqref{eq:bispectrum-covariance} (valid only in the weak non-Gaussianity regime. For visibility, $\ic_{\ell\ell'}$ is multiplied by the factor $\ell\ell'$. Note that the color scale is logarithmic, except for values smaller in absolute value than $10^{-26}$ where it becomes linear.}
    \label{fig:ib-covariance}  
\end{figure}

Similarly, binning the integrated bispectrum as we did for example in figure \ref{fig:ib_z1100} (mainly for readability reasons) has a limited effect on error bars. We have shown in section \ref{sec:simulations} that the diagonal part of the integrated bispectrum covariance matrix is well-approximated by eq.\ \eqref{eq:ibisp_variance}, as can be also checked in figure \ref{fig:binned-ibisp}, where we compare the theoretical covariance to the one estimated from the usual 40 simulations. In this figure, we also show that the error bar on the mean value of a bin of width 100 in multipole space is actually of the same order as the value from a single multipole $\ell$ (except at very low $\ell$), while it should be 10 times smaller if modes were uncorrelated. This confirms the large correlation between different integrated bispectrum modes. We can also go a step further with the integrated bispectrum and estimate its full covariance from simulations as shown in figure \ref{fig:ib-covariance} (left panel). This full covariance is compared to the theoretical counterpart, in the right panel, determined in the case of a diagonal full bispectrum covariance (eq.\ \ref{eq:bispectrum-covariance}). While the theoretical covariance in the weak NG limit is close to diagonal (non-zero terms are only found where $|\ell-\ell'| \leq 2\lwmax=20$), we can see that the actual covariance from simulations (left panel) is non-negligible everywhere. Even very different $\ell$ and $\ell'$ are correlated. This of course will need to be taken into account in more advanced applications, aimed for example at building a bispectrum/integrated bispectrum  likelihood to measure cosmological parameters.

A direct solution to characterize the full integrated bispectrum covariance matrix is simply to estimate it from many mock simulations, as exemplified in figure \ref{fig:binned-ibisp}. However, reaching the desired accuracy for actual parameter estimation and matrix inversion cannot be achieved with only 40 simulations, as done here for illustrative purposes. Considering the typical number of modes in our analysis, we will actually need thousands of simulations. Assuming these are available, a further issue is the time currently needed to extract the integrated bispectrum from a single map. We recall that most of the computational time is taken by the estimation of the power spectrum in each patch (192 power spectra per map in this paper). With our current approach and map resolution, the overall computational time becomes large, if we need to repeat the operation over many simulations (at $\lmax=2000$ each power spectrum takes $\sim 10$s to evaluate on a recent 8-core processor). Significant improvements over the current implementation of the estimator can however be achieved. Instead of the costly full-sky power spectrum estimator required with the step function patches used in this paper, one can in principle use much more localized patches based on needlets (see \cite{Jung:2020zne} for examples). With this choice, it is then possible compute power spectra using the flat-sky approximation in each small, localized patch (this should lead to large gains, both because the total number of pixels per patch is reduced by a factor $\sim 10^2$ and because of the improved computational scaling, allowed by FFT operations in flat sky approximation). These needlet patches are also azimuthally symmetric and hence theoretical expressions similar to eqs.\ \eqref{eq:ibisp-bisp} and \eqref{eq:ibisp_covariance} can easily be derived.

The alternative approach, namely deriving analytically the integrated bispectrum covariance matrix using eq.\ \eqref{eq:ibisp_covariance} presents several difficulties, like, on one side, the sheer number of terms in the sum (10 different multipole numbers) and, on the other, the costly evaluation of many 6$j$-symbols. On top of this, eq.\ \eqref{eq:ibisp_covariance} is also assuming that the full bispectrum covariance is already available and we only need to extract the integrated bispectrum covariance out of it. 
\section{Conclusion}
\label{sec:conclusion}

In this work, we have studied the full-sky integrated angular bispectrum of the weak lensing convergence field. The integrated bispectrum is a simple-to-compute statistic, which probes the squeezed limit of the full bispectrum, by measuring the large scale modulation of the field power spectrum from many localized sky patches. This study constitutes the first step towards building a full analysis pipeline for upcoming experiments, like \euclid. Our main goal is that of testing the accuracy of theoretical predictions and state-of-the-art simulations for the weak lensing convergence field. To this purpose, we find it useful to consider the largest scales in our analysis: on one hand, such scales are the easiest to model in a perturbative approach; on the other hand, theoretical models of the convergence field on the same scales are affected by the use of the Limber approximation, which introduces some intrinsic inaccuracy, which needs to be tested.

We therefore focus on measuring very squeezed bispectrum configurations ($\ell_1 \sim \ell_2 \gg \ell_3$, with $2 \leq \ell_3 \leq 10$), using the exact integrated bispectrum estimator on the full sky, originally developed in \cite{Jung:2020zne}. If we shift the focus purely on maximizing sensitivity, rather than on testing accuracy, then a higher signal-to-noise for the integrated signal can likely be achieved by picking less squeezed triangles ($\ell_3 \sim 100$), as the lensing signal is larger on smaller scales. During the reviewing process of this work, it was indeed shown in \cite{Halder:2021itp} that a fully flat-sky approach using many extremely small patches is also possible and measures a strong non-Gaussian signal without probing precisely the large-scale modes on which we focus in this paper. For the reasons just mentioned, we find the two approaches complementary and addressing different questions.

Our tests, based on weak-lensing convergence simulations at different source redshifts ($z_s=0.5,1,1.5,2$), show a good, but not perfect, agreement between the measured non-Gaussian signal and its theoretical prediction. This is the case both in an ideal situation (full-sky, noiseless maps) and with more realistic partial sky ($\fsky=0.35$) analyses, including Gaussian noise realizations. To check whether this mismatch could be due to some issues in our integrated bispectrum estimation approach, we have applied an independent method -- namely, the binned bispectrum estimator -- to the same set of simulations. We have thus verified that the two methods produce consistent results, displaying in both cases the same small mismatch between the measured and predicted non-Gaussian signal. 

We thus conclude that the mismatch arises from uncertainties on the theoretical, rather than on the numerical side of our pipeline. The theoretical bispectrum template was in fact already pointed out in previous works to be slightly less accurate on large scales. Indeed, using the binned bispectrum estimator, we have also explicitly shown that the observed discrepancy between theoretical predictions and numerical results is only present in the squeezed limit, where large scales do play an important role. 

Finally, we have explored the issue of how to precisely estimate the full integrated bispectrum covariance. 
This is an important point if we want to be able to use our integrated bispectrum pipeline for future applications, like cosmological parameter inference from weak lensing non-Gaussianity. In \cite{Jung:2020zne} we had shown that, thanks to our choice of azimuthally symmetric patches, the covariance could be quickly evaluated by means of a simple semi-analytical formula, valid in the weak non-Gaussianity limit of CMB analysis. However, such limit does not strictly apply to weak lensing and we have verified that this semi-analytical approach is no longer good enough for precise evaluation of the off-diagonal covariance terms (while it still holds quite well for the variance part). The obvious approach is then to evaluate the full covariance by Monte-Carlo averaging over many mock datasets, but this requires further developments to significantly speed up our current pipeline. We argue that such speed up is actually possible by exploiting strongly localized, azimuthally symmetric, needlet patches, and by resorting to the flat-sky approximation in the power spectrum estimation step of the method. At the same time, this will also make it possible to probe less squeezed configurations ($\ell_3 \sim 100$) where the signal-to-noise ratio is expected to be even larger. We leave this to future work, in which we will also investigate extensions of our method to compute the integrated angular bispectrum of the shear field. 
\\

{\bf Acknowledgements:} We would like to thank Peter Taylor for providing us his code to generate the Euclid-type mask used in our study. 

\noindent Some of the results in this paper have been derived using the healpy \cite{Zonca2019} and HEALPix packages.

\noindent GJ and ML were supported by the project "Combining Cosmic Microwave Background and Large Scale Structure data: an Integrated Approach for Addressing Fundamental Questions in Cosmology", funded by the MIUR Progetti di Ricerca di Rilevante Interesse Nazionale (PRIN) Bando 2017 - grant 2017YJYZAH. 

\noindent GJ, and ML also acknowledge support from the ASI-COSMOS Network (cosmosnet.it) and from the INDARK INFN Initiative (web.infn.it/CSN4/IS/Linea5/InDark), which provided access to CINECA supercomputing facilities (cineca.it)

\noindent DM is supported by a grant from the Leverhume Trust at MSSL.
\appendix

\section{Other binned bispectrum configurations}
\label{ap:binned-bispectrum-configurations}

\begin{figure}
    \centering
    \includegraphics[width=0.24\linewidth]{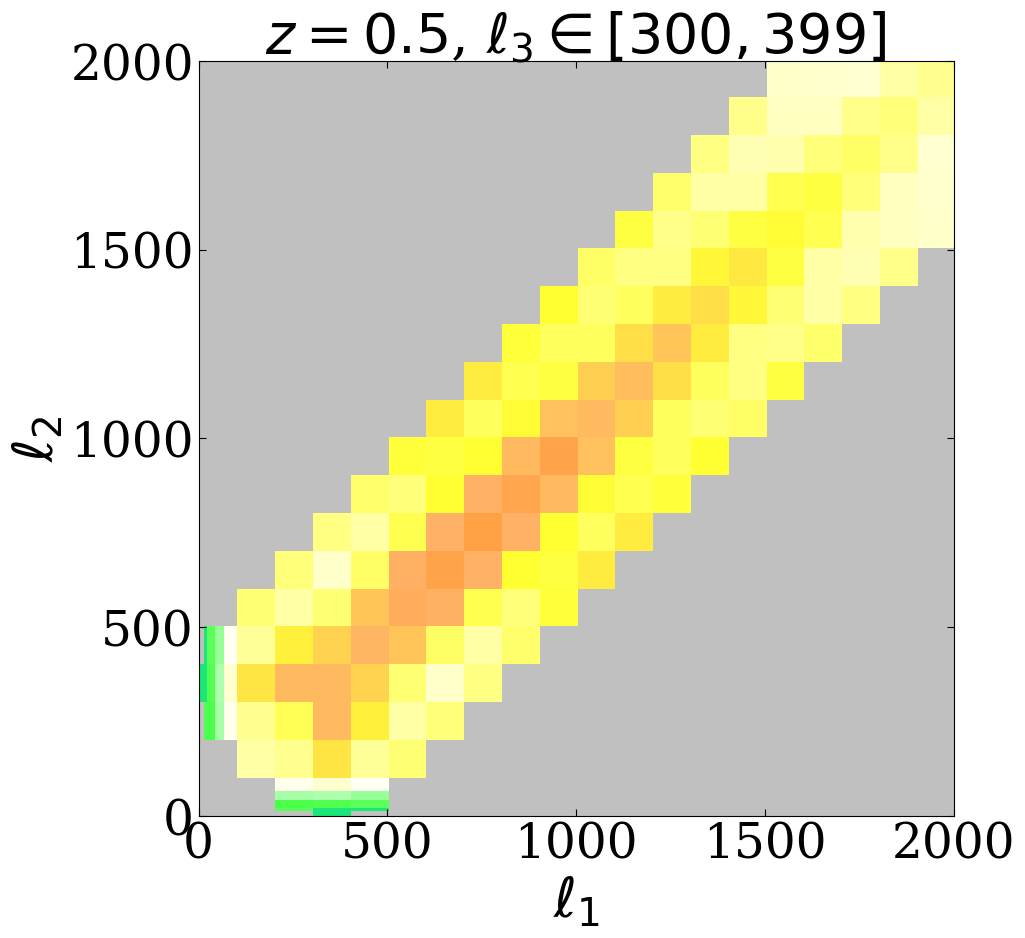}    
    \includegraphics[width=0.24\linewidth]{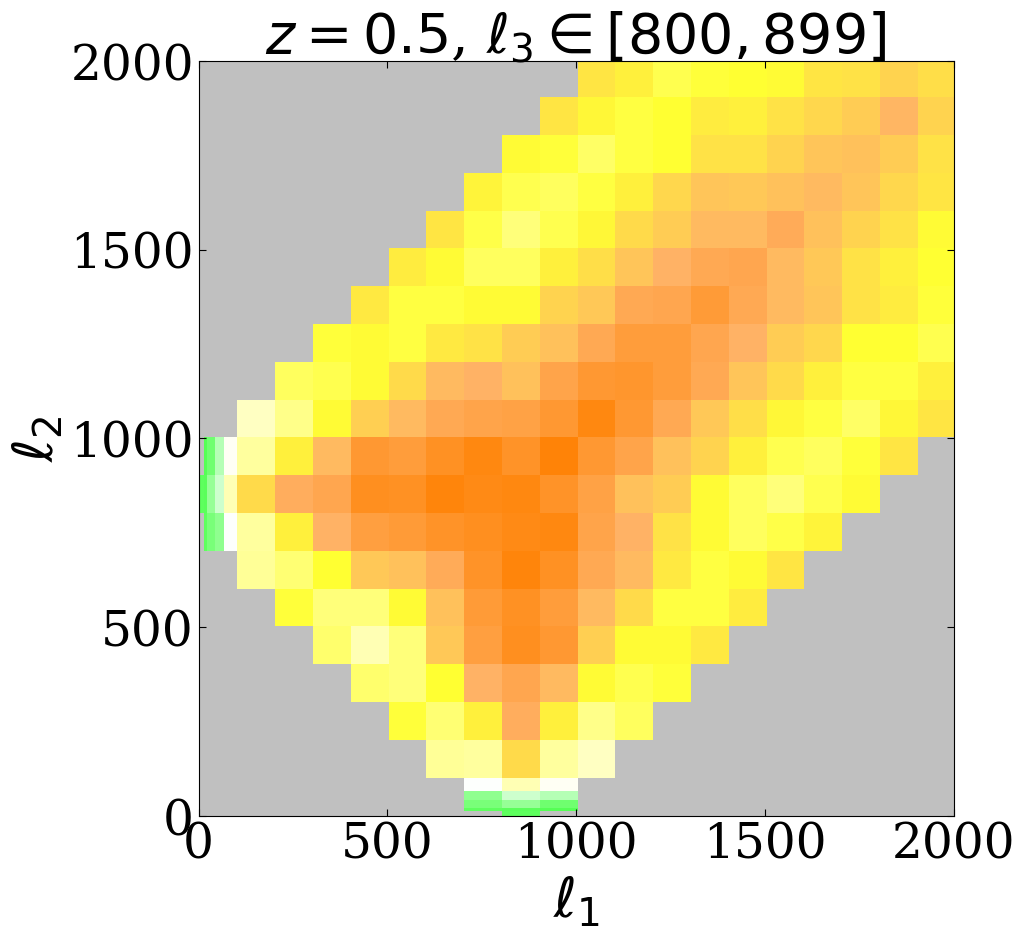}
    \includegraphics[width=0.24\linewidth]{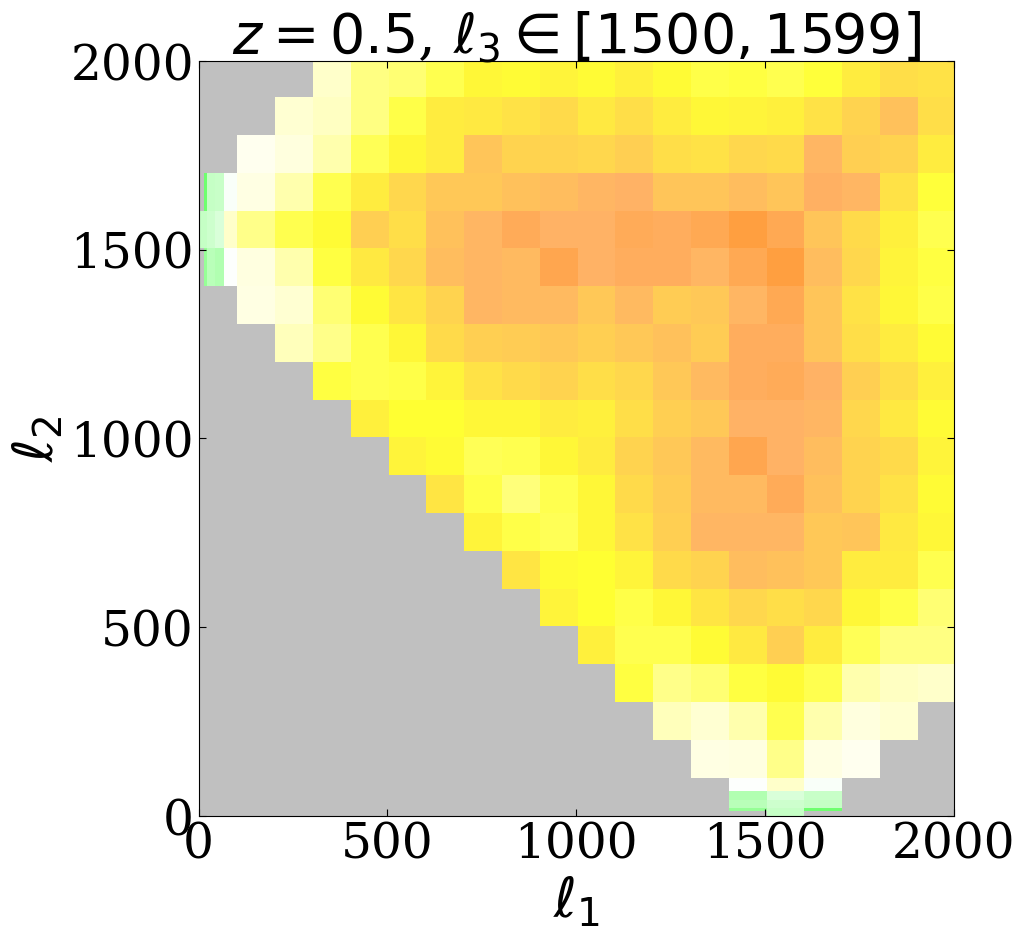}
    \includegraphics[width=0.24\linewidth]{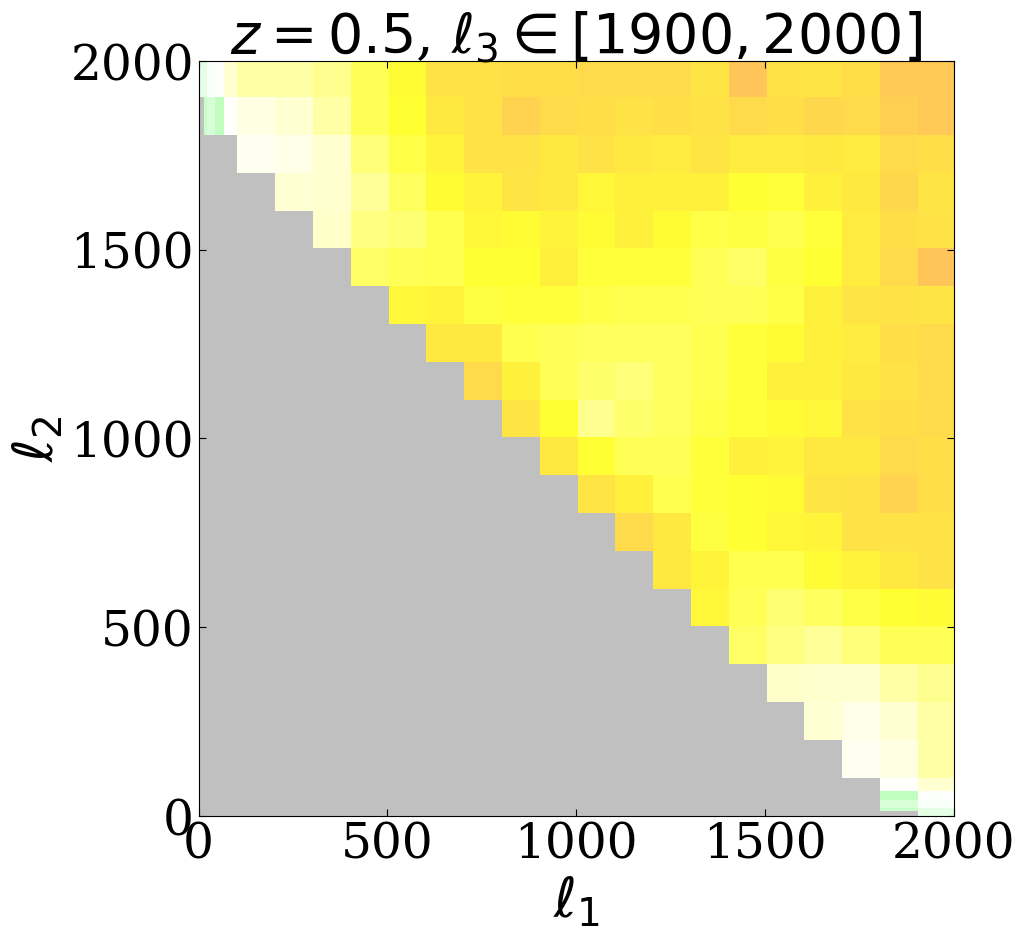}
    \includegraphics[width=0.24\linewidth]{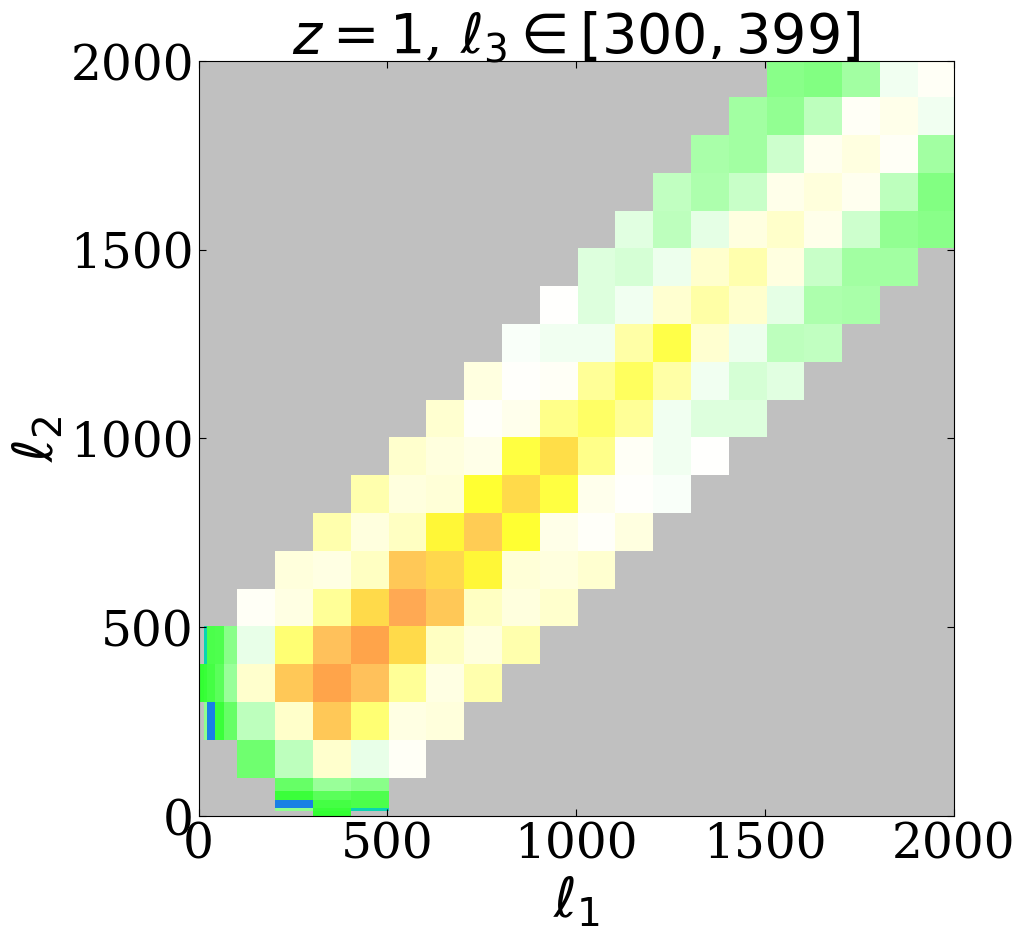}    
    \includegraphics[width=0.24\linewidth]{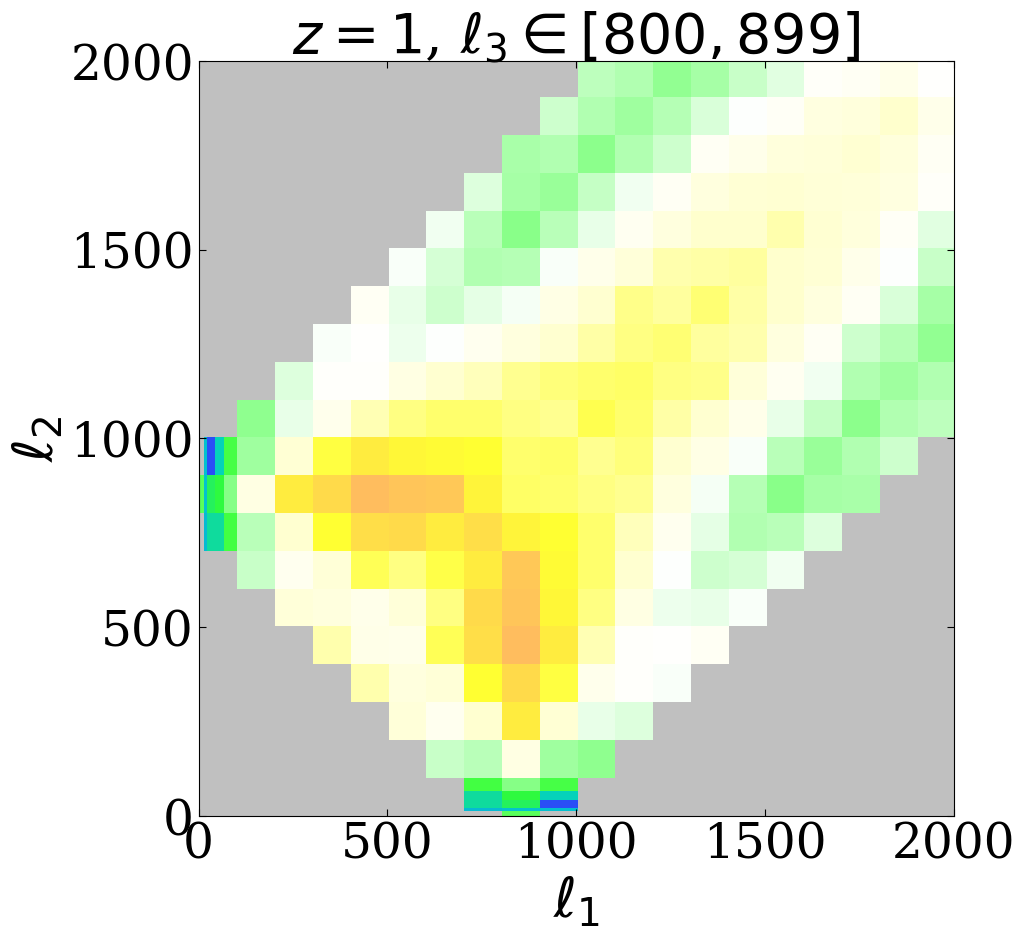}
    \includegraphics[width=0.24\linewidth]{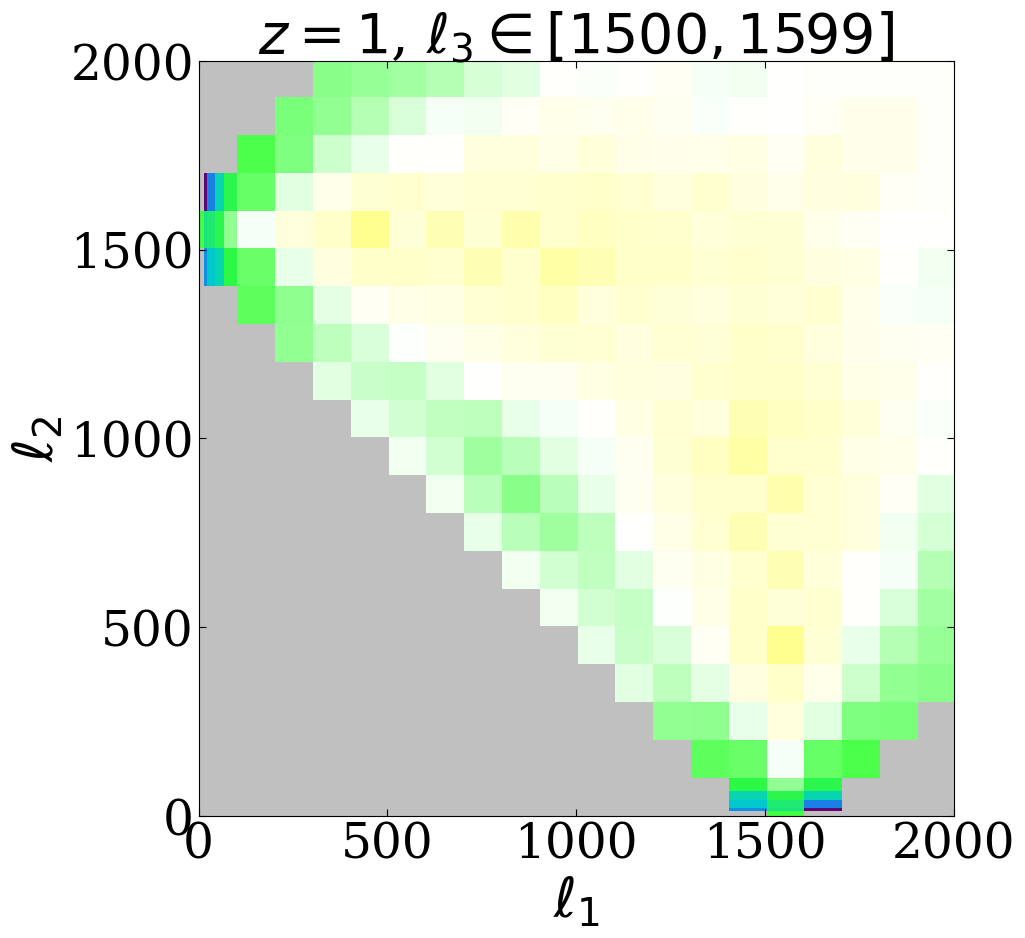}
    \includegraphics[width=0.24\linewidth]{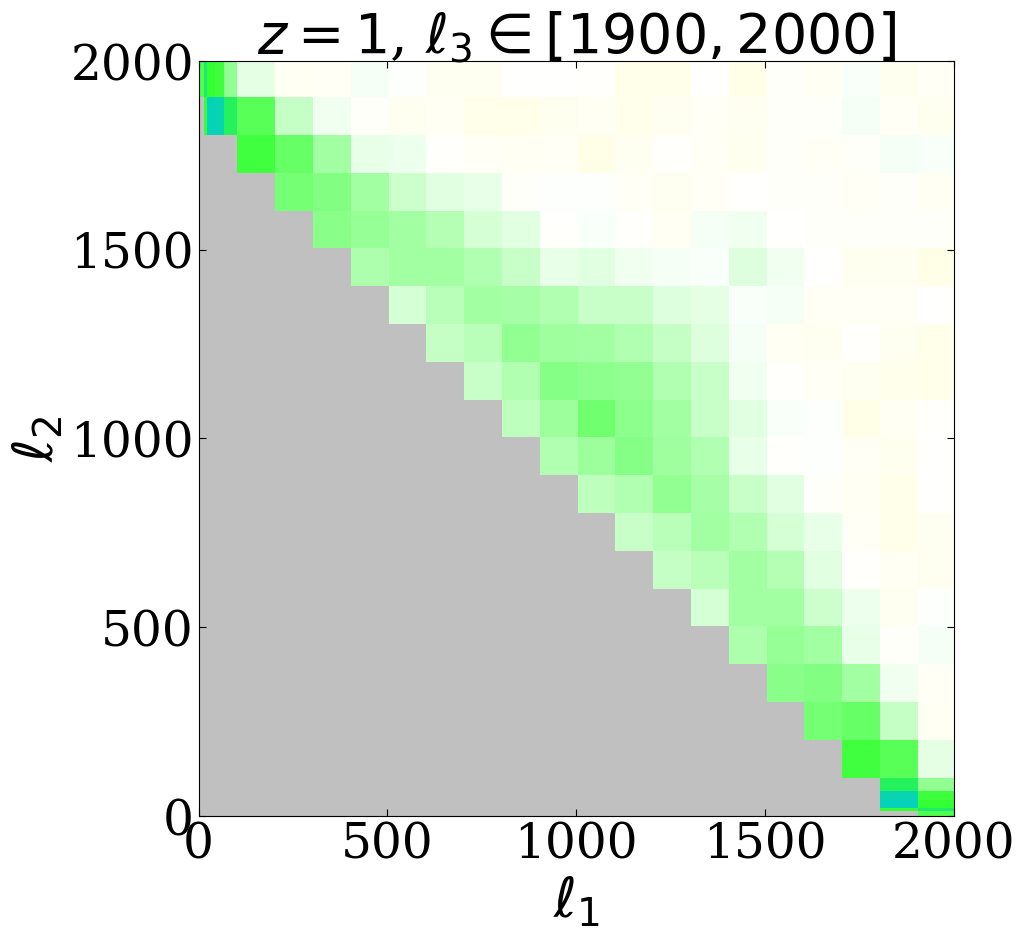}
    \includegraphics[width=0.24\linewidth]{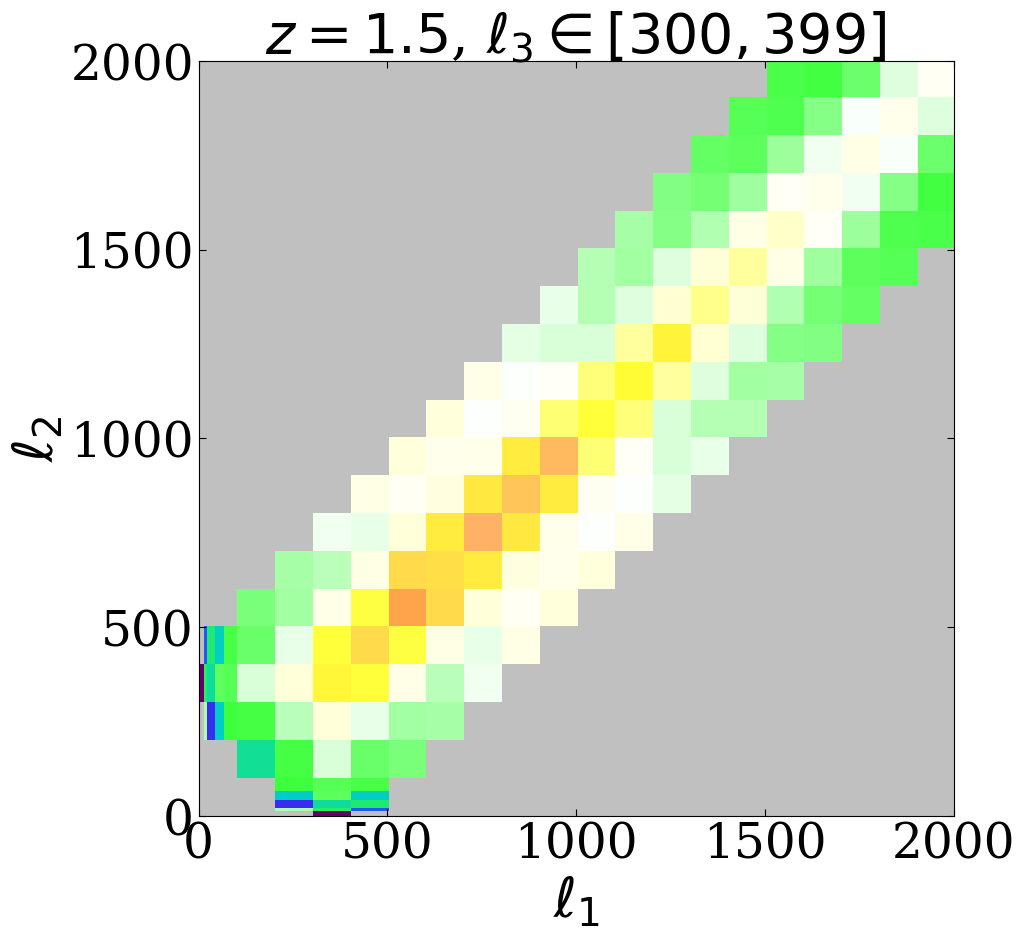}    
    \includegraphics[width=0.24\linewidth]{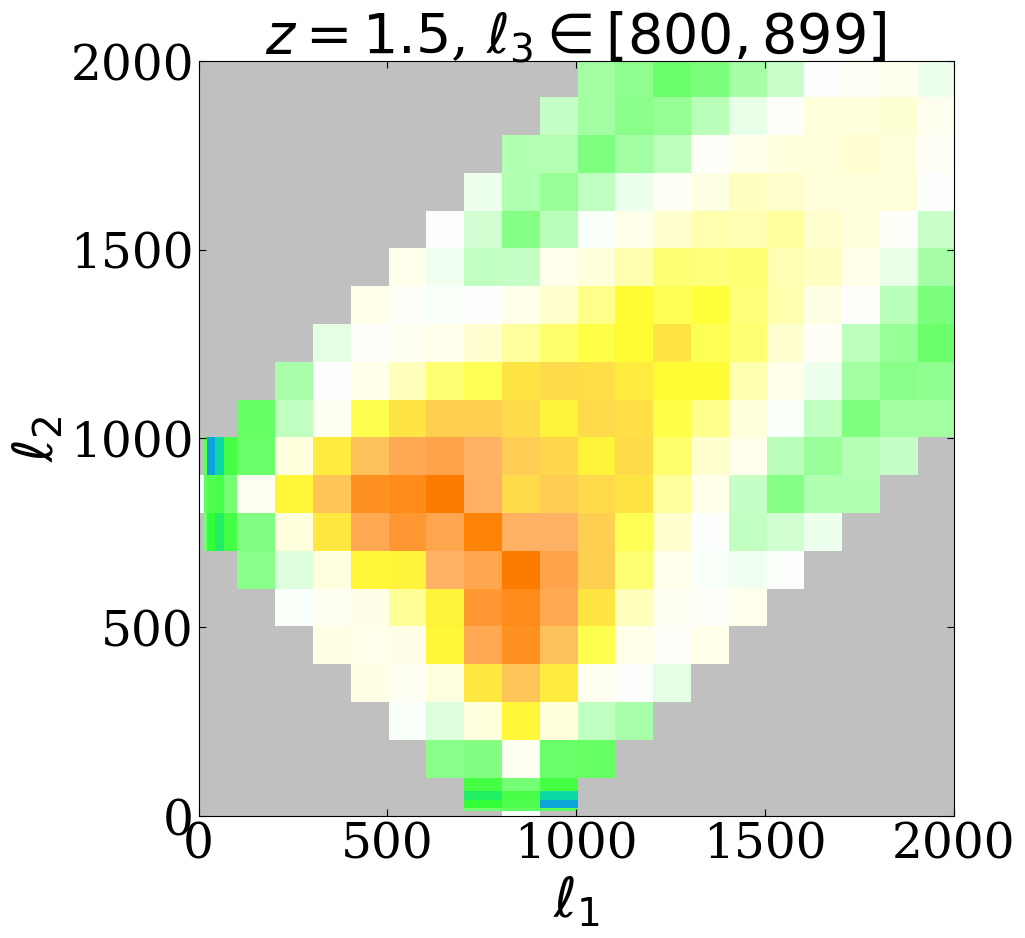}
    \includegraphics[width=0.24\linewidth]{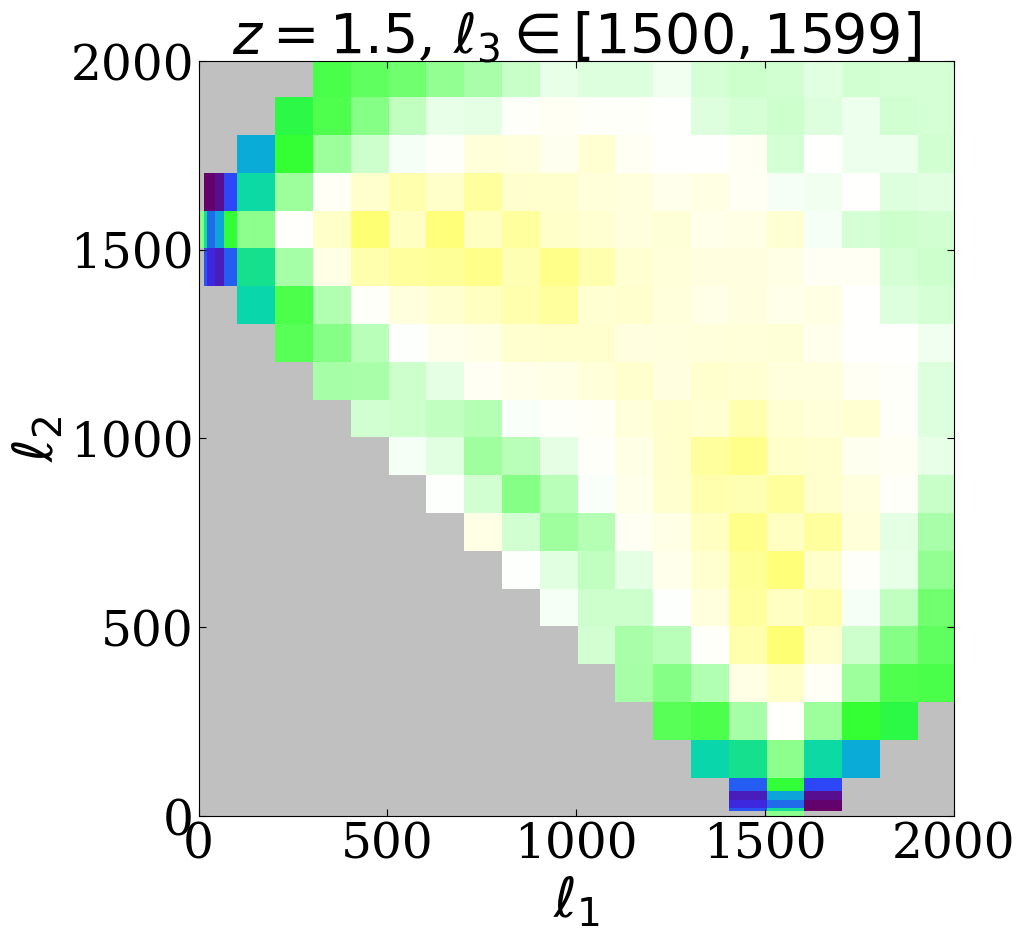}
    \includegraphics[width=0.24\linewidth]{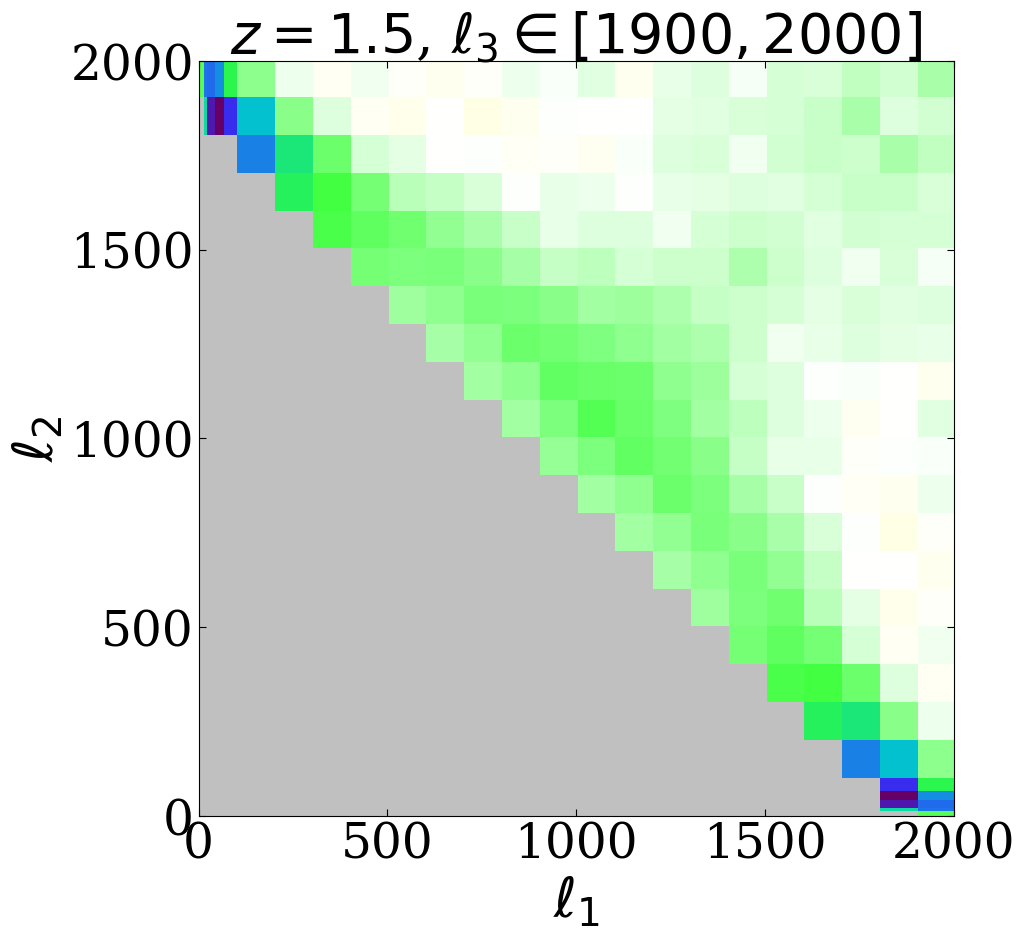}
    \includegraphics[width=0.24\linewidth]{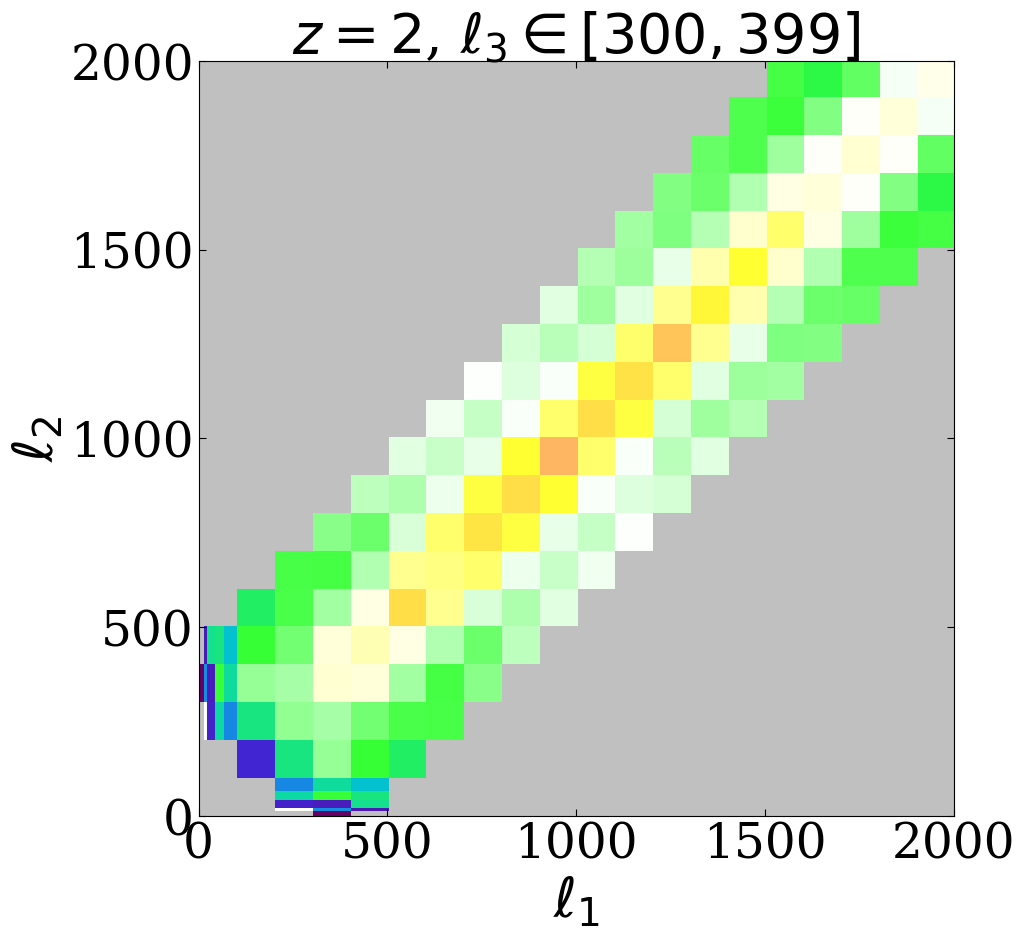}    
    \includegraphics[width=0.24\linewidth]{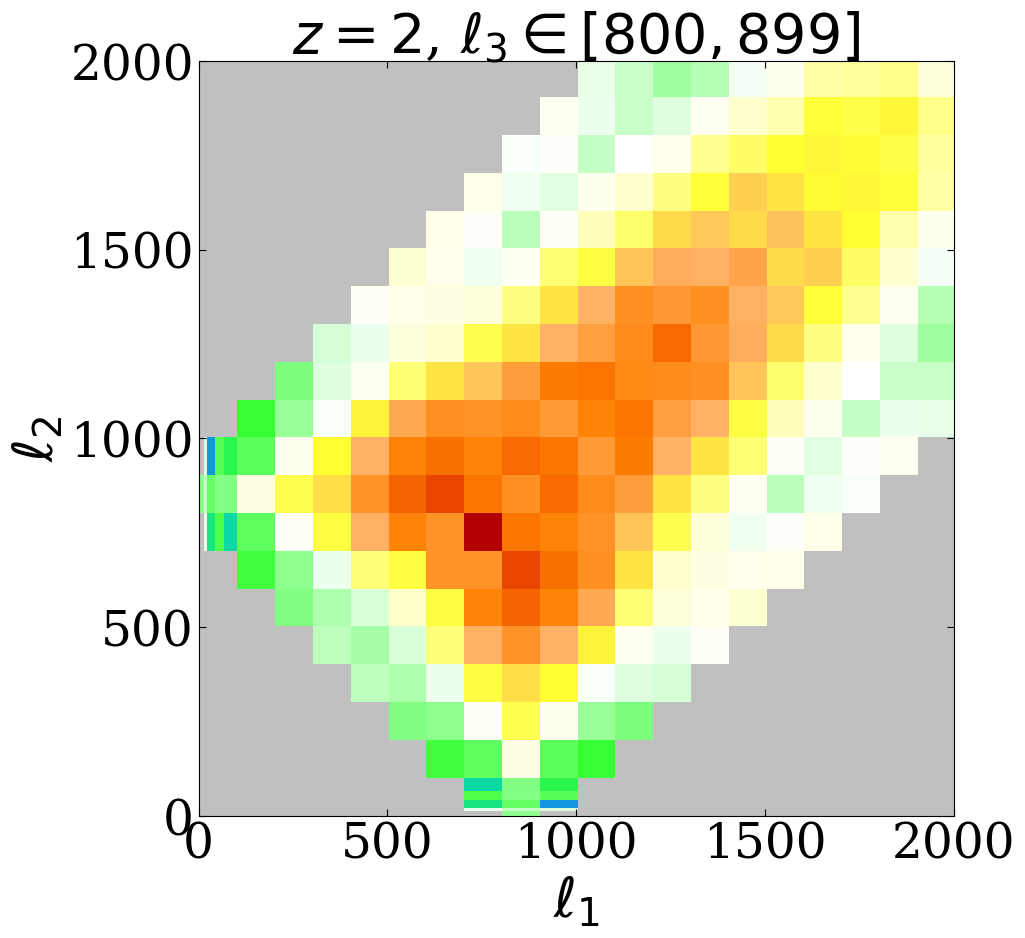}
    \includegraphics[width=0.24\linewidth]{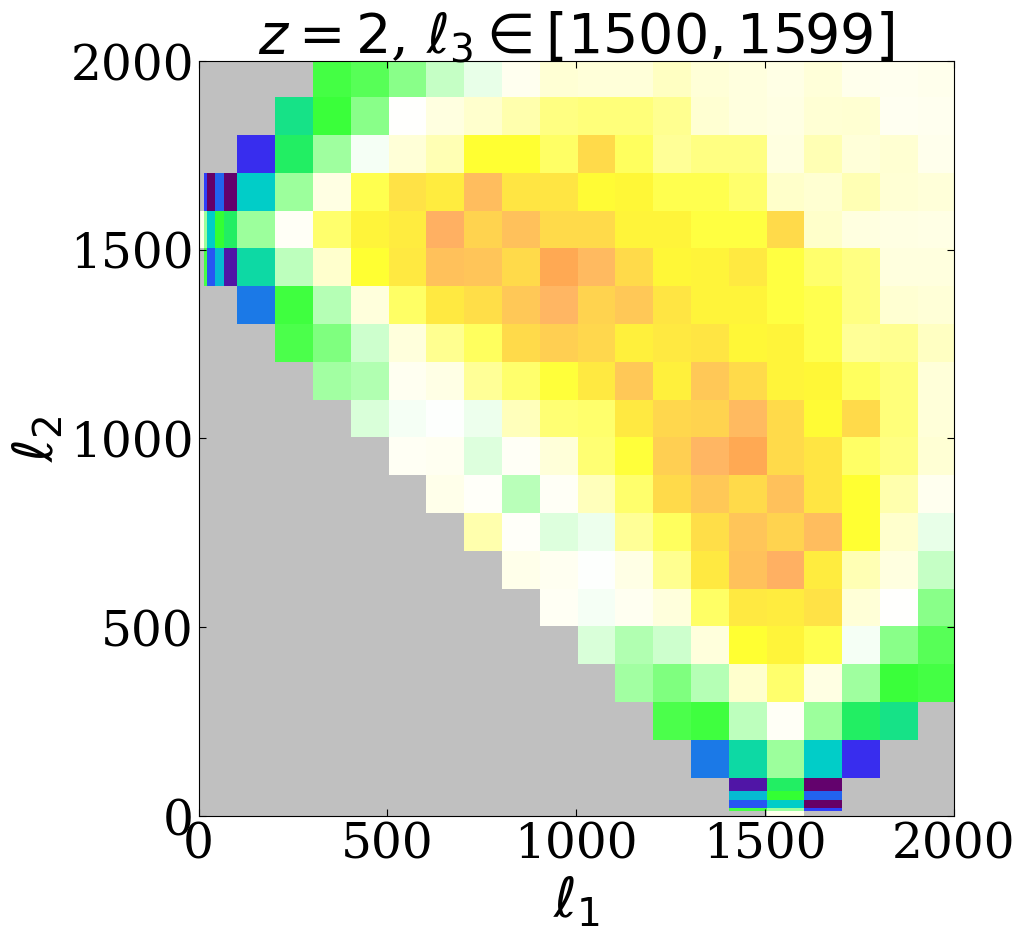}
    \includegraphics[width=0.24\linewidth]{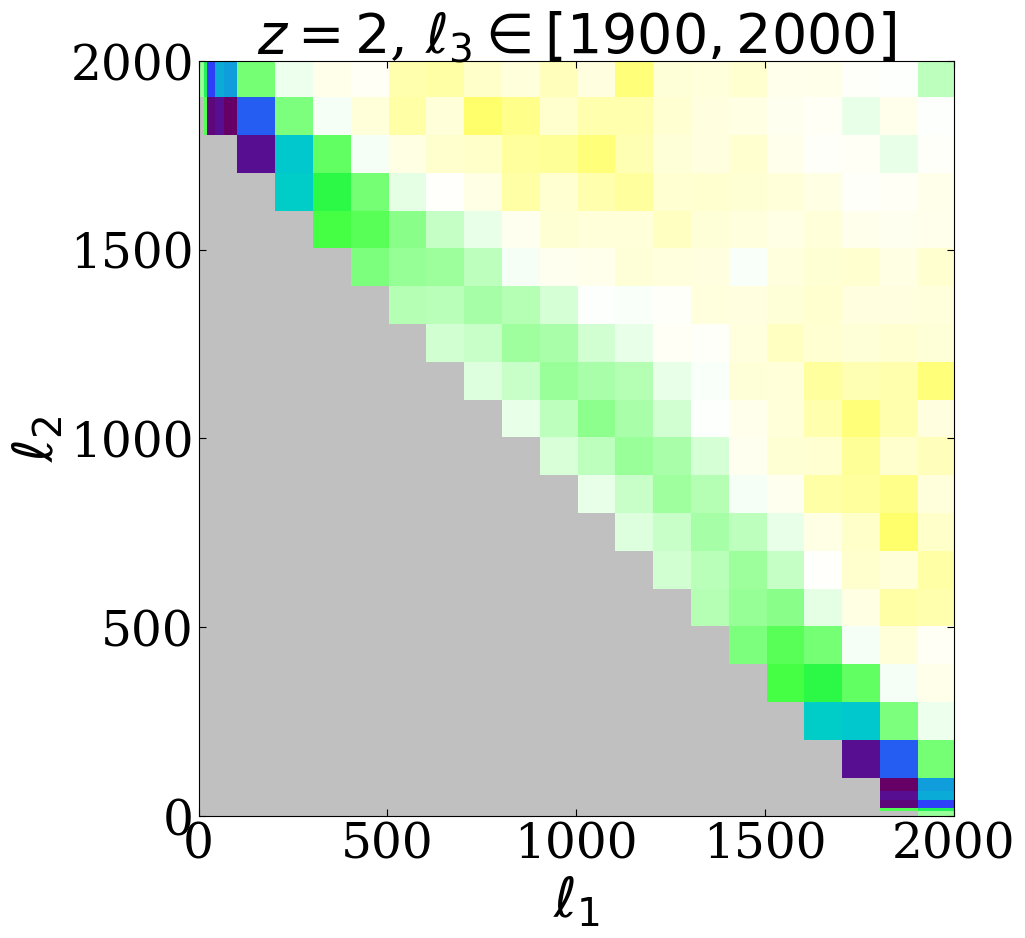}
    \includegraphics[width=0.4\linewidth]{figures/binned_bispectrum/colorbar_bispectrum_ratio.png}
    \caption{Same as figure \ref{fig:binbisp_4z} for different configurations.}
    \label{fig:binbisp}  
\end{figure}

In this appendix we follow the same binned bispectrum approach as in section \ref{sec:binned-bispectrum}, this time without focusing on the squeezed limit ($\ell_3 \ll \ell_1 \sim \ell_2$). In figure\ \ref{fig:binbisp}, we plot several ratios measured to predicted binned bispectra, similarly to figures \ref{fig:binbisp_4z} and \ref{fig:binbisp_zs1} but with larger $\ell_3$. In these plots, the squeezed limit can only be seen in the left and bottom corners ($\ell_1 \ll \ell_2 \sim \ell_3$ and $\ell_2 \ll \ell_1 \sim \ell_3$ respectively). This clearly shows that the theoretical templates are overestimated (ratio smaller than 1) mainly in the squeezed limit. Outside of these two corners, where other configurations are shown, the ratios are closer to 1 with a few exceptions. At $z_s=0.5$, where the non-Gaussianity is the largest and thus is more difficult to accurately modelize, the predicted bispectrum tends to be underestimated everywhere (except in the squeezed limit). At other redshifts, there is also a similar issue with close to equilateral configurations ($\ell_1 \sim \ell_2 \sim \ell_3$) with all $\ell$'s smaller 1000. The problem is the largest at $z_s=2$ when the weak lensing non-Gaussianity is the smallest meaning that the problem is different than at $z_s=0.5$.

\bibliographystyle{JHEP}
\bibliography{biblio}

\providecommand{\href}[2]{#2}\begingroup\raggedright\begin{thebibliography}{10}

\bibitem{Hikage2019}
C.~{Hikage}, M.~{Oguri}, T.~{Hamana}, S.~{More}, R.~{Mandelbaum}, M.~{Takada}
  et~al., \emph{{Cosmology from cosmic shear power spectra with Subaru Hyper
  Suprime-Cam first-year data}},
  \href{https://doi.org/10.1093/pasj/psz010}{\emph{Publ. Astr. Soc. Japan}
  {\bfseries 71} (2019) 43} [\href{https://arxiv.org/abs/1809.09148}{{\ttfamily
  1809.09148}}].

\bibitem{Troxel2019}
S.~{Samuroff}, J.~{Blazek}, M.~A. {Troxel}, N.~{MacCrann}, E.~{Krause}, C.~D.
  {Leonard} et~al., \emph{{Dark Energy Survey Year 1 results: constraints on
  intrinsic alignments and their colour dependence from galaxy clustering and
  weak lensing}}, \href{https://doi.org/10.1093/mnras/stz2197}{\emph{\mnras}
  {\bfseries 489} (2019) 5453}
  [\href{https://arxiv.org/abs/1811.06989}{{\ttfamily 1811.06989}}].

\bibitem{Asgari2020}
M.~{Asgari}, C.-A. {Lin}, B.~{Joachimi}, B.~{Giblin}, C.~{Heymans},
  H.~{Hildebrandt} et~al., \emph{{KiDS-1000 Cosmology: Cosmic shear constraints
  and comparison between two point statistics}}, {\emph{arXiv e-prints} (2020)
  arXiv:2007.15633} [\href{https://arxiv.org/abs/2007.15633}{{\ttfamily
  2007.15633}}].

\bibitem{Laureijs2011}
R.~{Laureijs}, J.~{Amiaux}, S.~{Arduini}, J.~L. {Augu{\`e}res},
  J.~{Brinchmann}, R.~{Cole} et~al., \emph{{Euclid Definition Study Report}},
  {\emph{arXiv e-prints} (2011) arXiv:1110.3193}
  [\href{https://arxiv.org/abs/1110.3193}{{\ttfamily 1110.3193}}].

\bibitem{Porqueres2021}
N.~{Porqueres}, A.~{Heavens}, D.~{Mortlock} and G.~{Lavaux}, \emph{{Bayesian
  forward modelling of cosmic shear data}}, {\emph{arXiv e-prints} (2020)
  arXiv:2011.07722} [\href{https://arxiv.org/abs/2011.07722}{{\ttfamily
  2011.07722}}].

\bibitem{Jeffrey2021}
N.~{Jeffrey}, J.~{Alsing} and F.~{Lanusse}, \emph{{Likelihood-free inference
  with neural compression of DES SV weak lensing map statistics}},
  \href{https://doi.org/10.1093/mnras/staa3594}{\emph{Mon. Not. Roy. Astron.
  Soc.} {\bfseries 501} (2021) 954}
  [\href{https://arxiv.org/abs/2009.08459}{{\ttfamily 2009.08459}}].

\bibitem{Takada2004}
M.~{Takada} and B.~{Jain}, \emph{{Cosmological parameters from lensing power
  spectrum and bispectrum tomography}},
  \href{https://doi.org/10.1111/j.1365-2966.2004.07410.x}{\emph{\mnras}
  {\bfseries 348} (2004) 897}
  [\href{https://arxiv.org/abs/astro-ph/0310125}{{\ttfamily
  astro-ph/0310125}}].

\bibitem{Kayo2013}
I.~{Kayo}, M.~{Takada} and B.~{Jain}, \emph{{Information content of weak
  lensing power spectrum and bispectrum: including the non-Gaussian error
  covariance matrix}},
  \href{https://doi.org/10.1093/mnras/sts340}{\emph{\mnras} {\bfseries 429}
  (2013) 344} [\href{https://arxiv.org/abs/1207.6322}{{\ttfamily 1207.6322}}].

\bibitem{Rizzato2019}
M.~{Rizzato}, K.~{Benabed}, F.~{Bernardeau} and F.~{Lacasa}, \emph{{Tomographic
  weak lensing bispectrum: a thorough analysis towards the next generation of
  galaxy surveys}}, \href{https://doi.org/10.1093/mnras/stz2862}{\emph{\mnras}
  {\bfseries 490} (2019) 4688}
  [\href{https://arxiv.org/abs/1812.07437}{{\ttfamily 1812.07437}}].

\bibitem{Pyne2020}
S.~{Pyne} and B.~{Joachimi}, \emph{{Self-calibration of weak lensing systematic
  effects using combined two- and three-point statistics}}, {\emph{arXiv
  e-prints} (2020) arXiv:2010.00614}
  [\href{https://arxiv.org/abs/2010.00614}{{\ttfamily 2010.00614}}].

\bibitem{Munshi2010}
D.~{Munshi} and A.~{Heavens}, \emph{{A new approach to probing primordial
  non-Gaussianity}},
  \href{https://doi.org/10.1111/j.1365-2966.2009.15820.x}{\emph{\mnras}
  {\bfseries 401} (2010) 2406}
  [\href{https://arxiv.org/abs/0904.4478}{{\ttfamily 0904.4478}}].

\bibitem{Munshi2020c}
D.~{Munshi}, T.~{Namikawa}, T.~D. {Kitching}, J.~D. {McEwen} and F.~R.
  {Bouchet}, \emph{{Weak lensing skew-spectrum}},
  \href{https://doi.org/10.1093/mnras/staa2769}{\emph{\mnras} {\bfseries 498}
  (2020) 6057} [\href{https://arxiv.org/abs/2006.12832}{{\ttfamily
  2006.12832}}].

\bibitem{Munshi2020b}
D.~{Munshi}, T.~{Namikawa}, J.~D. {McEwen}, T.~D. {Kitching} and F.~R.
  {Bouchet}, \emph{{Morphology of Weak Lensing Convergence Maps}}, {\emph{arXiv
  e-prints} (2020) arXiv:2010.05669}
  [\href{https://arxiv.org/abs/2010.05669}{{\ttfamily 2010.05669}}].

\bibitem{Bernardeau2012}
F.~{Bernardeau}, C.~{Bonvin}, N.~{Van de Rijt} and F.~{Vernizzi}, \emph{{Cosmic
  shear bispectrum from second-order perturbations in general relativity}},
  \href{https://doi.org/10.1103/PhysRevD.86.023001}{\emph{Phys. Rev. D}
  {\bfseries 86} (2012) 023001}
  [\href{https://arxiv.org/abs/1112.4430}{{\ttfamily 1112.4430}}].

\bibitem{Chiang:2014oga}
C.-T. Chiang, C.~Wagner, F.~Schmidt and E.~Komatsu, \emph{{Position-dependent
  power spectrum of the large-scale structure: a novel method to measure the
  squeezed-limit bispectrum}},
  \href{https://doi.org/10.1088/1475-7516/2014/05/048}{\emph{JCAP} {\bfseries
  1405} (2014) 048} [\href{https://arxiv.org/abs/1403.3411}{{\ttfamily
  1403.3411}}].

\bibitem{Chiang:2015eza}
C.-T. Chiang, C.~Wagner, A.~G. Sánchez, F.~Schmidt and E.~Komatsu,
  \emph{{Position-dependent correlation function from the SDSS-III Baryon
  Oscillation Spectroscopic Survey Data Release 10 CMASS Sample}},
  \href{https://doi.org/10.1088/1475-7516/2015/09/028,
  10.1088/1475-7516/2015/9/028}{\emph{JCAP} {\bfseries 1509} (2015) 028}
  [\href{https://arxiv.org/abs/1504.03322}{{\ttfamily 1504.03322}}].

\bibitem{Chiang:2015pwa}
C.-T. Chiang, \emph{{Position-dependent power spectrum: a new observable in the
  large-scale structure}}, Ph.D. thesis, Munich U., 2015.
\newblock \href{https://arxiv.org/abs/1508.03256}{{\ttfamily 1508.03256}}.

\bibitem{Giri:2018dln}
S.~K. Giri, A.~D'Aloisio, G.~Mellema, E.~Komatsu, R.~Ghara and S.~Majumdar,
  \emph{{Position-dependent power spectra of the 21-cm signal from the epoch of
  reionization}},
  \href{https://doi.org/10.1088/1475-7516/2019/02/058}{\emph{JCAP} {\bfseries
  1902} (2019) 058} [\href{https://arxiv.org/abs/1811.09633}{{\ttfamily
  1811.09633}}].

\bibitem{Munshi2017}
D.~{Munshi} and P.~{Coles}, \emph{{The integrated bispectrum and beyond}},
  \href{https://doi.org/10.1088/1475-7516/2017/02/010}{\emph{\jcap} {\bfseries
  2017} (2017) 010} [\href{https://arxiv.org/abs/1608.04345}{{\ttfamily
  1608.04345}}].

\bibitem{Munshi2020a}
D.~{Munshi}, J.~D. {McEwen}, T.~{Kitching}, P.~{Fosalba}, R.~{Teyssier} and
  J.~{Stadel}, \emph{{Estimating the integrated bispectrum from weak lensing
  maps}}, \href{https://doi.org/10.1088/1475-7516/2020/05/043}{\emph{\jcap}
  {\bfseries 2020} (2020) 043}
  [\href{https://arxiv.org/abs/1902.04877}{{\ttfamily 1902.04877}}].

\bibitem{Jung:2020zne}
G.~Jung, F.~Oppizzi, A.~Ravenni and M.~Liguori, \emph{{The integrated angular
  bispectrum}},
  \href{https://doi.org/10.1088/1475-7516/2020/06/035}{\emph{JCAP} {\bfseries
  06} (2020) 035} [\href{https://arxiv.org/abs/2004.03574}{{\ttfamily
  2004.03574}}].

\bibitem{Bucher:2009nm}
M.~Bucher, B.~Van~Tent and C.~S. Carvalho, \emph{{Detecting Bispectral Acoustic
  Oscillations from Inflation Using a New Flexible Estimator}},
  \href{https://doi.org/10.1111/j.1365-2966.2010.17089.x}{\emph{Mon. Not. Roy.
  Astron. Soc.} {\bfseries 407} (2010) 2193}
  [\href{https://arxiv.org/abs/0911.1642}{{\ttfamily 0911.1642}}].

\bibitem{Bucher:2015ura}
M.~Bucher, B.~Racine and B.~van Tent, \emph{{The binned bispectrum estimator:
  template-based and non-parametric CMB non-Gaussianity searches}},
  \href{https://doi.org/10.1088/1475-7516/2016/05/055}{\emph{JCAP} {\bfseries
  1605} (2016) 055} [\href{https://arxiv.org/abs/1509.08107}{{\ttfamily
  1509.08107}}].

\bibitem{Munshi:2019csw}
D.~Munshi, T.~Namikawa, T.~Kitching, J.~McEwen, R.~Takahashi, F.~Bouchet
  et~al., \emph{{The Weak Lensing Bispectrum Induced By Gravity}},
  \href{https://doi.org/10.1093/mnras/staa296}{\emph{Mon. Not. Roy. Astron.
  Soc.} {\bfseries 493} (2020) 3985}
  [\href{https://arxiv.org/abs/1910.04627}{{\ttfamily 1910.04627}}].

\bibitem{Alsing2015}
J.~{Alsing}, D.~{Kirk}, A.~{Heavens} and A.~H. {Jaffe}, \emph{{Weak lensing
  with sizes, magnitudes and shapes}},
  \href{https://doi.org/10.1093/mnras/stv1249}{\emph{Mon. Not. Roy. Astron.
  Soc.} {\bfseries 452} (2015) 1202}
  [\href{https://arxiv.org/abs/1410.7839}{{\ttfamily 1410.7839}}].

\bibitem{Duncan2014}
C.~A.~J. {Duncan}, B.~{Joachimi}, A.~F. {Heavens}, C.~{Heymans} and
  H.~{Hildebrandt}, \emph{{On the complementarity of galaxy clustering with
  cosmic shear and flux magnification}},
  \href{https://doi.org/10.1093/mnras/stt2060}{\emph{\mnras} {\bfseries 437}
  (2014) 2471} [\href{https://arxiv.org/abs/1306.6870}{{\ttfamily 1306.6870}}].

\bibitem{Duncan2016}
C.~A.~J. {Duncan}, C.~{Heymans}, A.~F. {Heavens} and B.~{Joachimi},
  \emph{{Cluster mass profile reconstruction with size and flux magnification
  on the HST STAGES survey}},
  \href{https://doi.org/10.1093/mnras/stw027}{\emph{Mon. Not. Roy. Astron.
  Soc.} {\bfseries 457} (2016) 764}
  [\href{https://arxiv.org/abs/1601.02023}{{\ttfamily 1601.02023}}].

\bibitem{Bartelmann2001}
M.~{Bartelmann} and P.~{Schneider}, \emph{{Weak gravitational lensing}},
  \href{https://doi.org/10.1016/S0370-1573(00)00082-X}{\emph{\physrep}
  {\bfseries 340} (2001) 291}
  [\href{https://arxiv.org/abs/astro-ph/9912508}{{\ttfamily
  astro-ph/9912508}}].

\bibitem{Munshi2008}
D.~{Munshi}, P.~{Valageas}, L.~{van Waerbeke} and A.~{Heavens},
  \emph{{Cosmology with weak lensing surveys}},
  \href{https://doi.org/10.1016/j.physrep.2008.02.003}{\emph{\physrep}
  {\bfseries 462} (2008) 67}
  [\href{https://arxiv.org/abs/astro-ph/0612667}{{\ttfamily
  astro-ph/0612667}}].

\bibitem{Kilbinger2015}
M.~{Kilbinger}, \emph{{Cosmology with cosmic shear observations: a review}},
  \href{https://doi.org/10.1088/0034-4885/78/8/086901}{\emph{Reports on
  Progress in Physics} {\bfseries 78} (2015) 086901}
  [\href{https://arxiv.org/abs/1411.0115}{{\ttfamily 1411.0115}}].

\bibitem{Takahashi:2019hth}
R.~Takahashi, T.~Nishimichi, T.~Namikawa, A.~Taruya, I.~Kayo, K.~Osato et~al.,
  \emph{{Fitting the nonlinear matter bispectrum by the Halofit approach}},
  \href{https://doi.org/10.3847/1538-4357/ab908d}{\emph{Astrophys. J.}
  {\bfseries 895} (2020) 113}
  [\href{https://arxiv.org/abs/1911.07886}{{\ttfamily 1911.07886}}].

\bibitem{Pratten:2016dsm}
G.~Pratten and A.~Lewis, \emph{{Impact of post-Born lensing on the CMB}},
  \href{https://doi.org/10.1088/1475-7516/2016/08/047}{\emph{JCAP} {\bfseries
  08} (2016) 047} [\href{https://arxiv.org/abs/1605.05662}{{\ttfamily
  1605.05662}}].

\bibitem{Gorski:2004by}
K.~Gorski, E.~Hivon, A.~Banday, B.~Wandelt, F.~Hansen, M.~Reinecke et~al.,
  \emph{{HEALPix - A Framework for high resolution discretization, and fast
  analysis of data distributed on the sphere}},
  \href{https://doi.org/10.1086/427976}{\emph{Astrophys. J.} {\bfseries 622}
  (2005) 759} [\href{https://arxiv.org/abs/astro-ph/0409513}{{\ttfamily
  astro-ph/0409513}}].

\bibitem{Creminelli:2005hu}
P.~Creminelli, A.~Nicolis, L.~Senatore, M.~Tegmark and M.~Zaldarriaga,
  \emph{{Limits on non-gaussianities from wmap data}},
  \href{https://doi.org/10.1088/1475-7516/2006/05/004}{\emph{JCAP} {\bfseries
  0605} (2006) 004} [\href{https://arxiv.org/abs/astro-ph/0509029}{{\ttfamily
  astro-ph/0509029}}].

\bibitem{Kayo_2012}
I.~Kayo, M.~Takada and B.~Jain, \emph{Information content of weak lensing power
  spectrum and bispectrum: including the non-gaussian error covariance matrix},
  \href{https://doi.org/10.1093/mnras/sts340}{\emph{Monthly Notices of the
  Royal Astronomical Society} {\bfseries 429} (2012) 344–371}.

\bibitem{Takahashi:2017hjr}
R.~Takahashi, T.~Hamana, M.~Shirasaki, T.~Namikawa, T.~Nishimichi, K.~Osato
  et~al., \emph{{Full-sky Gravitational Lensing Simulation for Large-area
  Galaxy Surveys and Cosmic Microwave Background Experiments}},
  \href{https://doi.org/10.3847/1538-4357/aa943d}{\emph{Astrophys. J.}
  {\bfseries 850} (2017) 24}
  [\href{https://arxiv.org/abs/1706.01472}{{\ttfamily 1706.01472}}].

\bibitem{Namikawa:2018bju}
T.~Namikawa, B.~Bose, F.~R. Bouchet, R.~Takahashi and A.~Taruya, \emph{{CMB
  lensing bispectrum: Assessing analytical predictions against full-sky lensing
  simulations}}, \href{https://doi.org/10.1103/PhysRevD.99.063511}{\emph{Phys.
  Rev. D} {\bfseries 99} (2019) 063511}
  [\href{https://arxiv.org/abs/1812.10635}{{\ttfamily 1812.10635}}].

\bibitem{Munshi:2020tzm}
D.~Munshi, T.~Namikawa, J.~McEwen, T.~Kitching and F.~Bouchet,
  \emph{{Morphology of Weak Lensing Convergence Maps}},
  \href{https://arxiv.org/abs/2010.05669}{{\ttfamily 2010.05669}}.

\bibitem{Hildebrandt2017}
H.~{Hildebrandt}, M.~{Viola}, C.~{Heymans}, S.~{Joudaki}, K.~{Kuijken},
  C.~{Blake} et~al., \emph{{KiDS-450: cosmological parameter constraints from
  tomographic weak gravitational lensing}},
  \href{https://doi.org/10.1093/mnras/stw2805}{\emph{\mnras} {\bfseries 465}
  (2017) 1454} [\href{https://arxiv.org/abs/1606.05338}{{\ttfamily
  1606.05338}}].

\bibitem{Akrami:2019izv}
{\scshape Planck} collaboration, \emph{{Planck 2018 results. IX. Constraints on
  primordial non-Gaussianity}},
  \href{https://doi.org/10.1051/0004-6361/201935891}{\emph{Astron. Astrophys.}
  {\bfseries 641} (2020) A9}
  [\href{https://arxiv.org/abs/1905.05697}{{\ttfamily 1905.05697}}].

\bibitem{Halder:2021itp}
A.~Halder, O.~Friedrich, S.~Seitz and T.~N. Varga, \emph{{The integrated
  3-point correlation function of cosmic shear}},
  \href{https://arxiv.org/abs/2102.10177}{{\ttfamily 2102.10177}}.

\bibitem{Zonca2019}
A.~Zonca, L.~Singer, D.~Lenz, M.~Reinecke, C.~Rosset, E.~Hivon et~al.,
  \emph{healpy: equal area pixelization and spherical harmonics transforms for
  data on the sphere in python},
  \href{https://doi.org/10.21105/joss.01298}{\emph{Journal of Open Source
  Software} {\bfseries 4} (2019) 1298}.

\end{thebibliography}\endgroup

\end{document}